\pdfoutput=1
\documentclass{osa-article}

\journal{oe}

\usepackage{subfig}

\articletype{Research Article}

\begin{document}

\title{Introducing Coherent MIMO Sensing, a fading-resilient, polarization-independent approach to $\phi$-OTDR}

\author{Sterenn Guerrier,\authormark{1,*} Christian Dorize,\authormark{1}  Elie Awwad,\authormark{2} and J\'{e}r\'{e}mie Renaudier\authormark{1} }

\address{\authormark{1}Nokia Bell Labs France Paris-Saclay, route de Villejust, 91620 Nozay, France\\
\authormark{2}Department of Communications and Electronics, TELECOM Paris, 19 place Marguerite Perey, 91120 Palaiseau, France\\

}

\email{\authormark{*}sterenn.guerrier@nokia-bell-labs.com} 



\begin{abstract}
Nowadays, long distance optical fibre transmission systems use polarization diversity multiplexed signals to enhance transmission performance. 
Distributed acoustic sensors (DAS) use the same propagation medium ie. single mode optical fibre, and aims at comparable targets such as covering the highest distance with the best signal quality. 
In the case of sensors, a noiseless transmission enables to monitor a large quantity of mechanical events along the fibre. 
This paper aims at extending the perspectives of DAS systems with regard to technology breakthroughs introduced in long haul transmission systems over the last decade. 
We recently developed a sensor interrogation method based on coherent phase-sensitive optical time domain reflectometry ($\phi$-OTDR), with dual polarization multiplexing at the transmitter and polarization diversity at the receiver.
We name this technique Coherent-MIMO sensing. 
A study is performed from a dual-polarization numerical model to compare several sensor interrogation techniques, including Coherent-MIMO. 
We demonstrate that dual-polarization probing of a fibre sensor makes it insensitive to polarization effects, decreases the risks of false alarms and thus strongly enhances its sensitivity.
The simulations results are validated with an experiment, and finally quantitative data are given on the performance increase enabled by Coherent-MIMO sensing. 
\end{abstract}


\section{Introduction} 
Optical fibres, the ones dedicated for sensing applications as much as the ones in already deployed cables in telecom infrastructure, can be employed as distributed acoustic and strain sensors. 
These sensors can serve in many fields, from marine environment monitoring \cite{Hartog2018_advances} to medical applications \cite{beisenova_fiber-optic_2018}. 
Often chosen because they are independent of most external physical fields (electromagnetic field impacts copper cables, temperature of the fibre strongly modifies the intensity in Raman fibre sensors) \cite{palmieri2013_distributed}, Rayleigh scattering-based sensors exploit the intensity, polarization and phase of light. 
When a disturbance occurs in the neighbourhood of the fibre, the induced intensity variations (intensity-based optical time domain reflectometry - OTDR) or phase variations (differential phase OTDR) of the backscattered light are captured to localize the disturbance source. 
The last years have seen the development of coherent $\phi$-OTDR, capable of distributed strain and temperature sensing \cite{pastor_2016_temperature}
over long distances and with tunable, up to sub-meter spatial resolution \cite{Lu_2017_highSpaRes}.
Coherent $\phi$-OTDR exploits the retro-propagated optical field induced by the Rayleigh backscattering effect in optical fibres, by measuring the backscattered phase of the fibre sensor. 
The main advantage of phase over intensity is that it offers a linear response to strain \cite{dorize_enhancing_2018,SEAFOM_2018}. 
It can also capture rapid changes in the disturbances, over high bandwidth. 
Rayleigh scattering however, is known to be polarization-dependent \cite{Deventer1993}. Besides, coherent setups are vulnerable to coherent noise also known as interference fading \cite{goodman_fundamental_1976,healey_fading_1984}, not to mention laser-related noise \cite{Fleyer2015} which will be out of the scope of this paper.
Coherent fading noise can be suppressed using specific pulse-probing methods \cite{chen_phase-detection_2017}.
To overcome polarization fading effect, recent work focus on the receiver design and uses a dual-polarization coherent receiver \cite{yan_coherent_2017,martins_real_2016} so that the overall received optical power is kept constant independently from the state of polarization of the incident signal at the receiver side. 
A wide spread assumption is that polarization-diversity receivers are the key to mitigate polarization fading \cite{yang_guangyao_polarization_2016}. Still, polarization-diversity reception can be improved \cite{gu_comparison_2018}. 
What's more, the dual-polarization receiver setup is not sufficient to provide a perfect channel estimation. 
The transmitter setup must as well be taken into account for fully achieving a polarization insensitive sensor.
Actually, polarization induced phase noise is transmitter-dependent,
which was observed in \cite{Kersey_1988_Observation} and further analysed in \cite{Kersey_1990_Analysis}. This effect depends on the input state of polarization (SOP). 
In the following, we compare single and dual polarization interrogation schemes, considering Single polarization Input - Single polarization Output (SISO), Single Input - Multiple Output ($1\times 2$ SIMO) and Multiple Input - Multiple Output ($2\times 2$ MIMO) probing techniques, and investigate the issues related to each of them.
A numerical model is developed to investigate the separate and joint effects of physical aspects encountered along an optical fibre. 
The transmitter, sensed fibre, and receiver are modelled as in \cite{Guerrier2019_Model}. Finally, we demonstrate improved sensitivity of $2\times 2$ MIMO, and call this sensing technique Coherent-MIMO Sensing. 

The paper is organized as follows. Section~\ref{sec:backscatterModel} details the numerical model used for the simulations, under static conditions and with the inclusion of noise contributions. Section~\ref{sec:phaseest} investigates the different manners to recover and compute the phase from the received backscattered field, and summarizes the main fading effects that can be identified and mitigated with modern sensing schemes. 
Section~\ref{sec:multiple_comp} compares the performance of interrogation methods when the number of used polarization channels varies using simulations; in this section, polarization diversity is used at the receiver side, simulations are run to identify and discuss the remaining sensitivity issues. 
Finally, section~\ref{sec:expres} details the experimental setup that validates simulations results and assesses the sensitivity of the developed MIMO sensor. 

\section{Construction of a dual-polarization Rayleigh backscatter fibre model}\label{sec:backscatterModel}
\subsection{Dual-polarization fibre model}\label{subsec:fibermodel} 
Several existing models investigated the backscattered phase and amplitude from optical fibres. 
The first step is to consider the fibre under static conditions \cite{Liokumovich2015}, the second is to add perturbations to the simulated fibre \cite{masoudi_analysis_2017}. 
We restrict ourselves to the static approach and introduce the polarization dimension to the model: the channel is seen as $\mathrm{N}$ successive fibre segments of length $L_s$, and each of these segments will be associated not only to the returned phase and amplitude (complex scalar), but to its complete Jones $2\times 2$ matrix \cite{Guerrier2019_Model}, denoted $\textbf{H}$, to also cover the polarization state dimension. 
Our scope is long SSMF (standard single mode fibres), with a highly coherent source (fibre length $L_f$ very small compared to source coherence length $L_{coh}$). 
A global phasor $p_i$ is derived for each segment $i \in [1,\mathrm{N}]$, which includes the local Rayleigh backscattering amplitude. 
In addition, $A_i$ the dual-pass attenuation in the fibre applied to segment $i$, is defined as $e^{2L_ia}$ where $L_i$ and $a$ are fibre distance from interrogator to the segment $i$ and linear attenuation in the fibre respectively, such that the received backscattered amplitude and phase from fibre segment $i$ will be $A_i p_i$.
\begin{figure}[htb]
\centering
\includegraphics[scale=0.55]{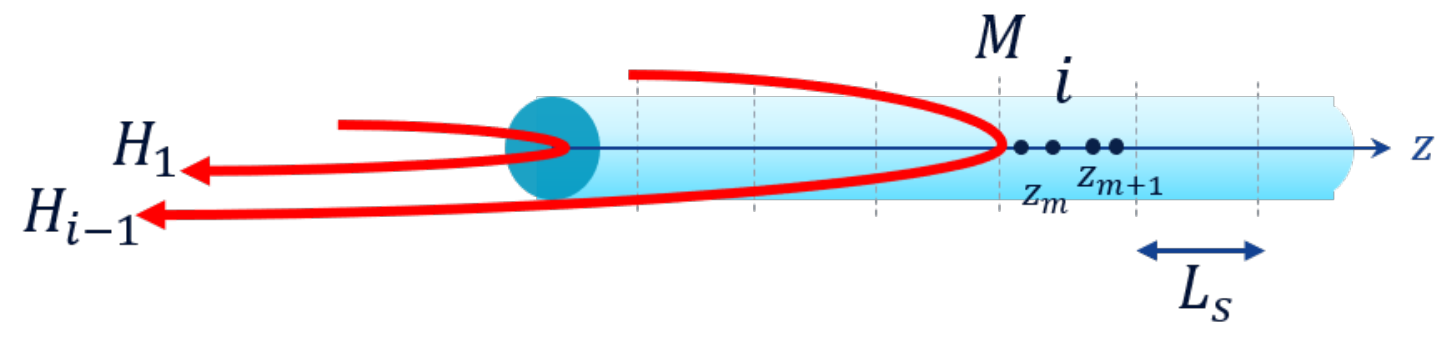}
\caption{Rayleigh scatterers modelling. Segment index is $i$, scatterer index within segment $i$ is $m$. $\textbf{H}_i$ is the Jones $2\times 2$ matrix describing segment $i$.}\label{fig:fiberschema}
\end{figure}\\
Extending our scope to dual-polarization, the phase and amplitude are distributed over two orthogonal polarization axes in the fibre. 
To each segment $i \in [1,\mathrm{N}]$, a single-pass, unitary Jones matrix $\textbf{U}_i$ is associated. 
It comprises the random polarization effects encountered by the light travelling through the fibre segment (forward transmission). 
To model these polarization effects, three random unitary matrices are generated: $\textbf{D}_ {\beta_i} = 
\mathrm{diag}(e^{j\beta_i} ; e^{-j\beta_i}) $  
and $\textbf{D}_ {\gamma_i} = 
\mathrm{diag}(e^{j\gamma_i} ; e^{-j\gamma_i}) $,  
diagonal phase retarders with $\gamma_i$ and $\beta_i$ uniformly drawn in $[-\pi, \pi]$ ($x \mapsto e^x$ is the exponential function), and $\textbf{R}_{\Theta_i} = 
 [\begin{smallmatrix}
\cos{\Theta_i} && -\sin{\Theta_i} \\ \sin{\Theta_i} && \cos{\Theta_i} \end{smallmatrix}]$ 
a polarization rotation real matrix, with $\Theta_i = \arcsin{\sqrt{\xi_i }}$ and $\xi_i$ uniformly drawn in $[0,1]$ \cite{BOYA_2003_volumes}. $\textbf{U}_i$ is the following product: 
\begin{equation}
\textbf{U}_i = e^{j\phi} \textbf{D}_ {\beta_i} \textbf{R}_{\Theta_i} \textbf{D}_{\gamma_i}
\label{eq:Ui} \end{equation}
where $\phi$ is the common phase term to cover all possible unitary matrices. It results from Eq.~\eqref{eq:Ui} that $\textbf{U}_i$ is a random unitary matrix,
which is accurate to describe the polarization state of any static fibre segment \cite{Damask2004_chap2}. 
In the following, as the interest is on the differential phase between segments, the common phase $\phi$ won't be considered in the expression of $\textbf{U}$. 
Generated this way, Jones matrices cover all possible states of polarization (SOP). 
For verification, it is possible to switch from Jones space to Stokes vectors and project the simulated SOP on the Poincar\'{e} sphere: for a significant number of random drawings, the full unit-radius sphere is covered.  \\
As we are interested in the backscattered light, the forward and reflection paths are modelled for each fibre segment. 
The reflection is placed at the end of the fibre segment. 
It can be modelled either as a perfect reflection $\textbf{M}= [\begin{smallmatrix} 1 && 0 \\ 0 && 1 \end{smallmatrix}]$
, or considering polarization transfers at the reflection, introducing a transfer coefficient $\alpha \in [0,0.05]$: 
$\textbf{M}_\alpha= [\begin{smallmatrix} \sqrt{1-\alpha} & -\sqrt{\alpha} \\ \sqrt{\alpha} & \sqrt{1-\alpha} \end{smallmatrix}]$.  
The value of this polarization transfer coefficient has to be measured and verified. 
The existence of such a transfer coefficient was considered for physical properties of silica \cite{sosman1927properties}, independently of optical fibre transmission. 
It could result in slight polarization changes of the Rayleigh backscattered light \cite{Hartog2017_chap2}, which we estimate below 5\% of light transferred from one polarization state to its orthogonal state. 
As it is not considered in most studies on Rayleigh backscatter, $\alpha = 0$ is set in the following, if not specified otherwise. 

The optical fibre is reciprocal from a propagation point of view, which 
translates into
$\textbf{U}_{backward} = \textbf{U}_{forward}^T$, with $T$ standing for the transpose operator \cite{Ross1982}\footnote{We change the formalism for the round-trip Jones matrix with respect to the one we used in \cite{Guerrier2019_Model,Awwad2020_JLT} to align with \cite{Ross1982} which better describes the SOP evolution after a round-trip in an optical fibre for each particular realization of $\mathbf{U_{forward}}(\beta,\Theta,\gamma)$ 
(Indeed, if we are only interested in phase noise statistics averaged over several realizations of backscattered SOP, both models are statistically equivalent. However, for each particular realization, the new model better depicts the SOP evolution).}.
We assume that the forward SOP is expressed in an (x,y,z) coordinate system where the z axis points towards the propagation axis and the backward SOP is expressed in the (x,y,-z) coordinate system. 
Assuming that $\textbf{U}_i$ can be the forward matrix of segment $i$ as well as the product of all forward matrices from segment 1 to segment $i$ (since a product of random unitary matrices is a random unitary matrix),
the full dual-pass Jones matrix from the initial segment up to fibre segment $i$ is then: 
\begin{equation}
\textbf{H}_i = A_i p_i \textbf{U}_i^T \textbf{M} \textbf{U}_i
\end{equation}  
where $A_i$ and $p_i$ stand for the dual-pass attenuation and phasor from the fibre start, and where $|p_i|$ follows a Rayleigh distribution \cite{Liokumovich2015}. 
Polarization dependent loss (PDL) and polarization mode dispersion (PMD) are considered negligible in the fibre sensor. 
The general shape of the Jones backscatter matrix 
is computed as follows for all spatial segments $i$: 
 \begin{equation*}
\textbf{H}_i = A_i p_{i} \begin{bmatrix}
e^{j2\gamma_i}( \cos2\beta_i + j\sin 2\beta_i \cos 2\Theta_i) &&
-j\sin 2\beta_i \sin 2\Theta_i \\
-j\sin 2\beta_i \sin 2\Theta_i &&
e^{-j2\gamma_i}( \cos2\beta_i - j\sin 2\beta_i \cos 2\Theta_i)
\end{bmatrix}\
\end{equation*} 
As the SOP of the emitted signal at the transmitter and the incident signal at the receiver side aren't perfectly aligned, a misalignment parameter $\theta$ will be considered. 
It is reported as an additional rotation, for example at the entrance of the system $\textbf{R}_{\theta} = [\begin{smallmatrix}
\cos{\theta} && -\sin{\theta} \\ \sin{\theta} && \cos{\theta} \end{smallmatrix}]$, resulting in
 $ \textbf{H}_{i} = A_i p_i \textbf{U}_i^T \textbf{M} \textbf{U}_i \textbf{R}_{\theta}$. 
No phase retarder is added as the considered fibre portions at the transmitter and receiver are small compared to the fibre sensor length $L_f$. 

Finally, the polarization beat length $L_{pb}$ of the simulated fibre is considered. 
This way, the speed of polarization variations from one fibre segment to the other is controlled.
For single mode fibres, $L_{pb}$ is of the order of magnitude of 10m \cite{corsi_beat_1999}.
By definition if $L_s \geq L_{pb}$, then the SOP in two successive segments is uncorrelated: 
polarization parameters are randomly drawn from one segment to the other, the SOP of  light returning from segment $i+1$ is independent from the SOP of light returning from segment $i$.
Else if $L_s/L_{pb} \rightarrow 0 $, then there is almost no SOP change from one fibre segment to the next. 
For values of $L_s/L_{pb}$ comprised in $]0,1[$, the change in the SOP will be proportional to the ratio $L_s/L_{pb}$, 
with $\lambda_{i+1} = \lambda_i + \frac{L_s}{L_{pb}}\times \lambda_{New}$, with $i$ the segment index,  $\lambda$ any polarization parameter (referring to $\beta$, $\gamma$ or $\Theta$), $\lambda_{New}$ randomly drawn as described in the beginning of this section and independent from $\lambda_i$. 

\begin{figure}[hbtp]
\centering
\captionsetup[subfigure]{oneside,margin={0cm,0cm}}
\subfloat[$L_s/L_{pb} = 1.4.10^{-2}$]{\includegraphics[width=0.33\textwidth]{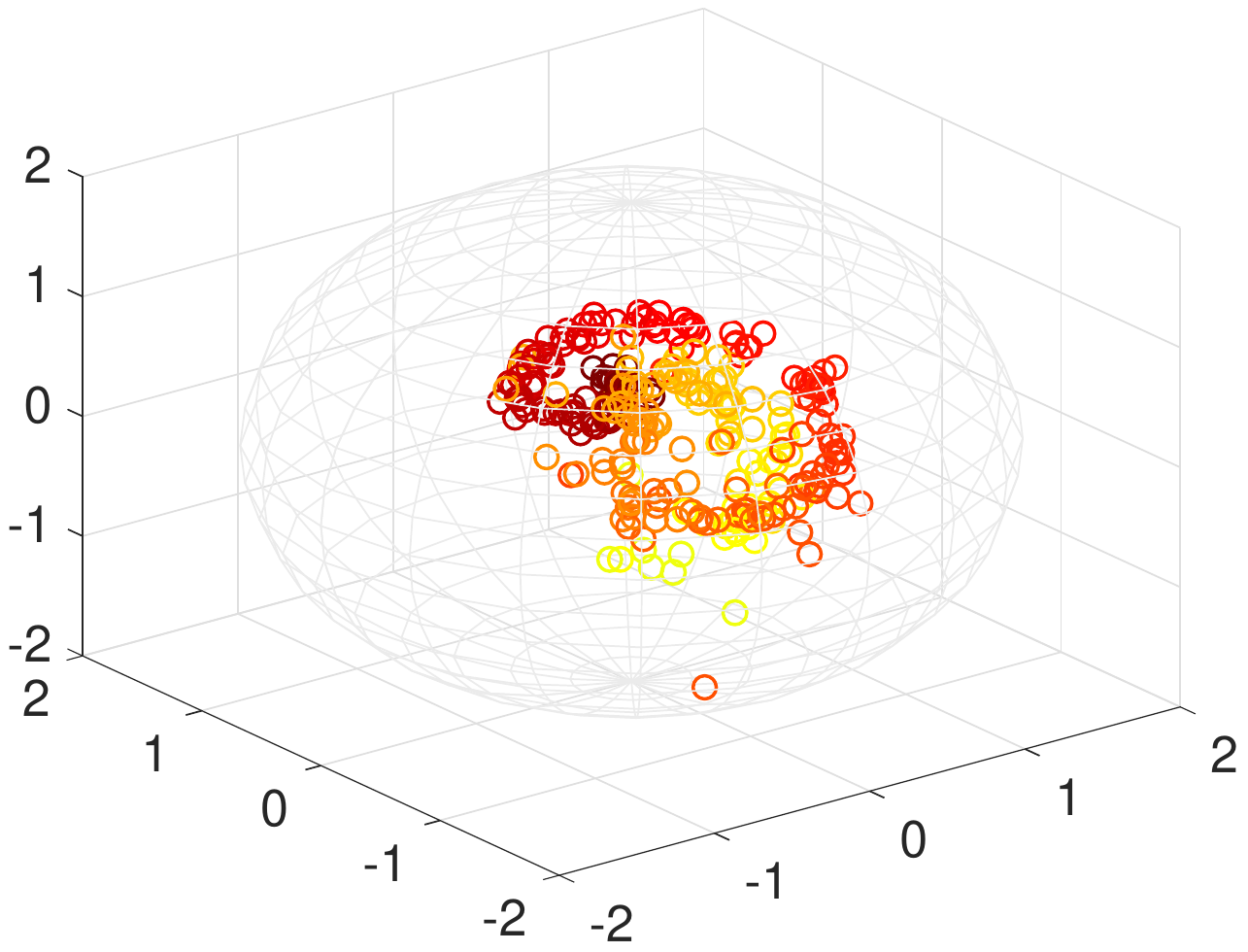}}
\subfloat[$L_s/L_{pb} = 6.8.10^{-2}$]{\includegraphics[width=0.33\textwidth]{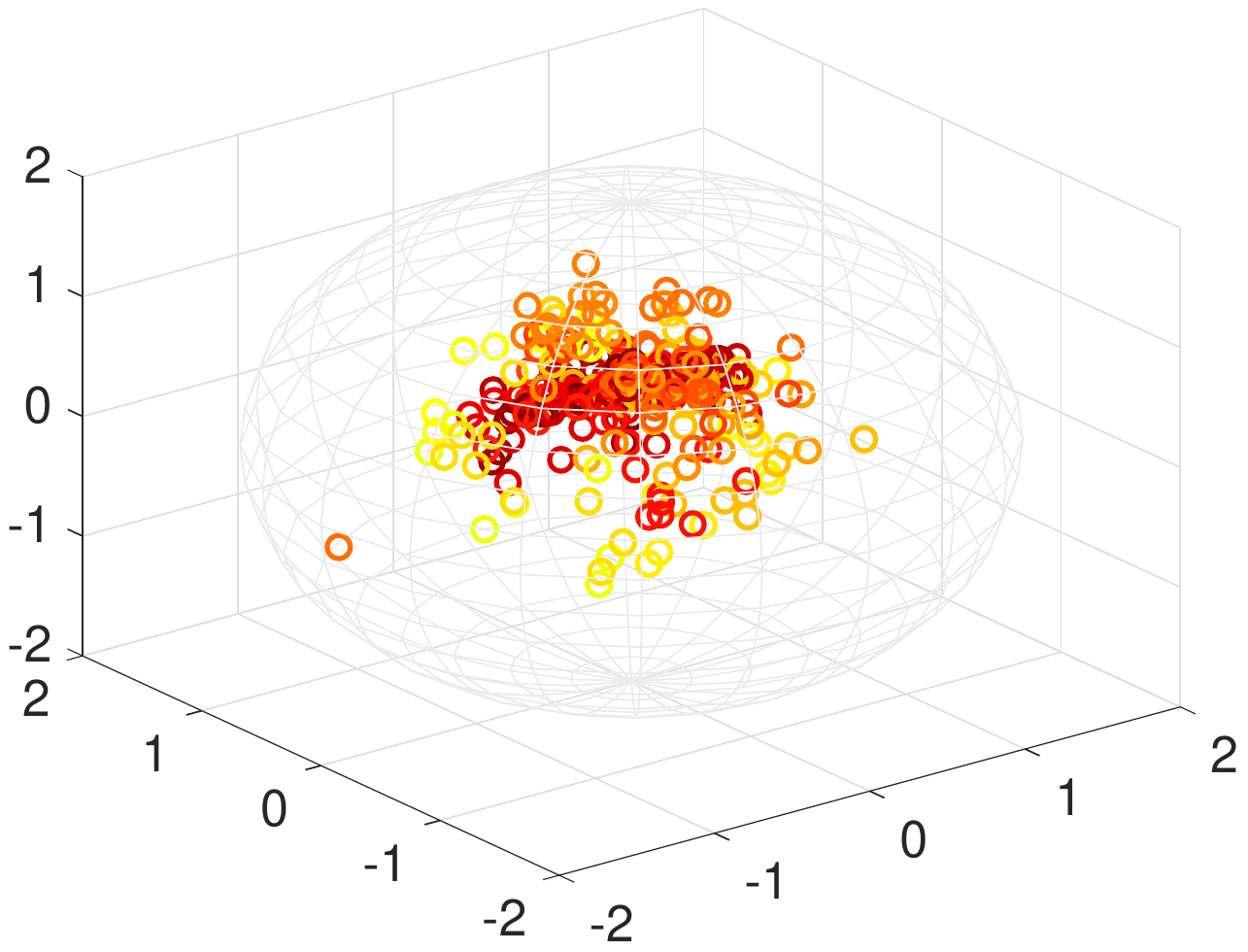}}
\subfloat[$L_s/L_{pb} = 3.4.10^{-1}$\label{subfig:lb2}]{\includegraphics[width=0.33\textwidth]{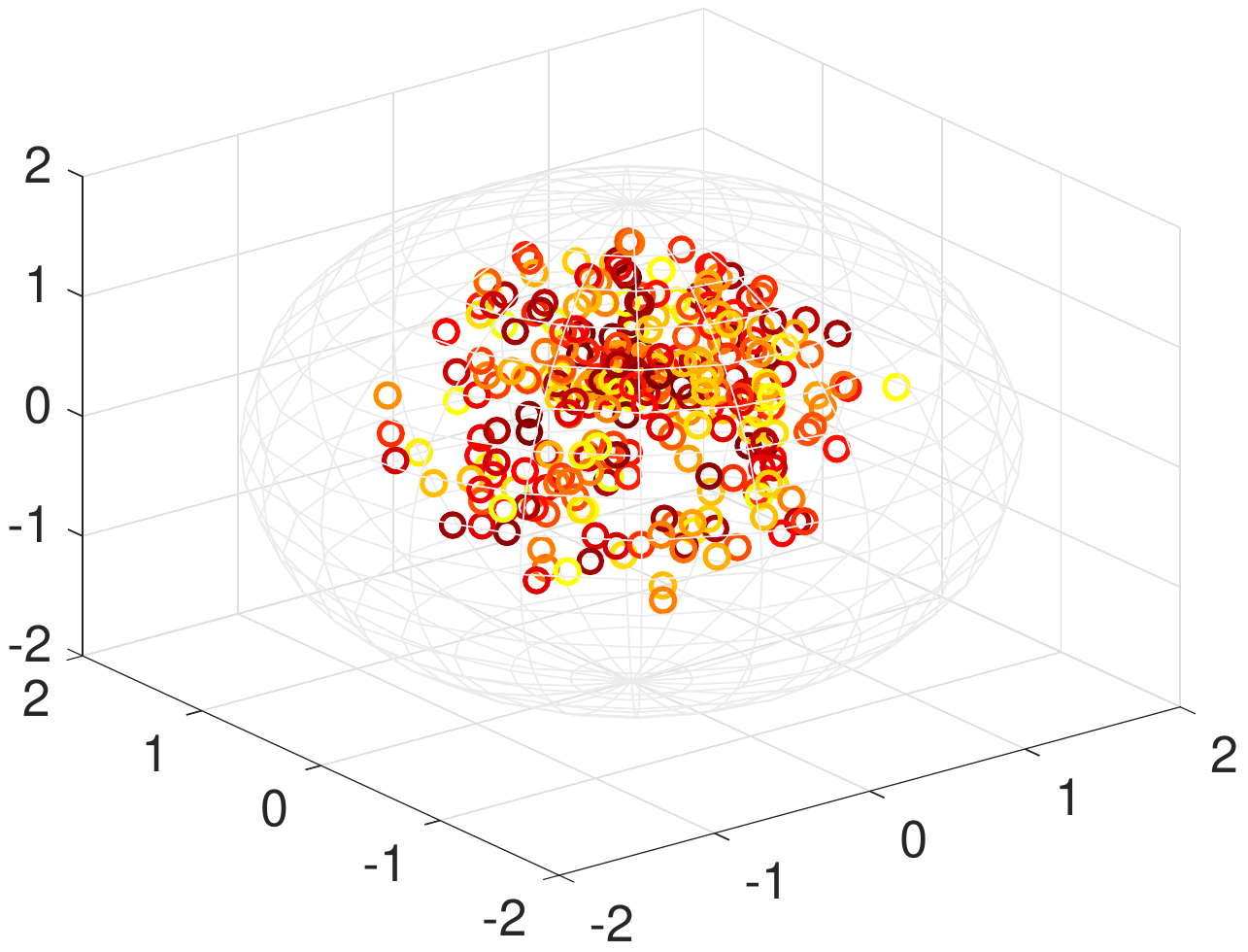}}
\caption{Poincar\'{e} sphere of polarization evolution along the fibre  (forward transmission), 
for different spatial resolutions $L_s$. Color changes with distance to fibre start.}\label{fig:poincareLbsmooth}
\vspace{-7pt}
\end{figure}

Several simulations for different $L_s/L_{pb}$ ratios are shown in \autoref{fig:poincareLbsmooth}, where it appears that the larger $L_s/L_{pb}$, the more random the trajectory of the SOP of light in the fibre (including attenuation $A_i$, thus not of constant modulus on the sphere). 
\\

It appears in \autoref{fig:poincareLbsmooth}\subref{subfig:lb2} that from $10 L_s > L_{pb}$, the distribution of the SOP over the Poincar\'{e} sphere is already random. 
When targeting sensing applications over telecom fibres, the spatial resolution $L_s$ is typically in the range of 1m to 10m (specific fields such as medicine are out of the scope of our study). 
Therefore, and for sake of simplicity, the simulation study below is conducted with uncorrelated SOP from one segment to the next ($L_s \geq L_{pb}$ case). 
Notice that additional simulations performed with $L_s \simeq 10^{-1} L_{pb}$ and  $L_s \geq L_{pb}$ in the sensor, which are not displayed here, lead to similar results. 

\subsection{DAS system model}
Similar to \cite{masoudi_analysis_2017}, our sensing model is considered as an ensemble of building blocks, so that elements can be added and removed at will to test their respective influence on measurement accuracy, noise, or phase retrieval. 
The transmitter encloses a narrow linewidth laser, modulated by a specific pattern.
Laser phase noise here is modelled as a Wiener process of variance $\sigma^2=2\pi\Delta\nu T_S$ where $\Delta\nu$ is the laser linewidth, $T_S$ is the sampling period. 
Although this does not perfectly match the physical reality, 
for which laser frequency noise increases in $1/f^d$, $d \geq 1$  \cite{Fleyer2015}, this enables to predict the general evolution of the error on phase estimation over distance \cite{Awwad2020_JLT}. 
The propagation medium is the fibre, the model of which is described in the previous section. 
The receiver is made of a coherent mixer, subject to insertion losses and additive white noise caused by thermal and quantum noise from the photodiodes.
 The white noise of the receiver is considered as $\sigma_{AWGN}^2 = \sigma_{th}^2 + \sigma_{s}^2 + \sigma_{RIN}^2$, where the contributions of the shot noise $\sigma_{s}^2$ and thermal noise $\sigma_{th}^2$ are dominant compared to the relative intensity noise (RIN) in our case. 
Altogether, we previously showed \cite{Guerrier2019_Model} that the laser phase noise is an important limitation in such a system. 
\begin{figure}[htbp]
\centering
\includegraphics[width=0.87\textwidth]{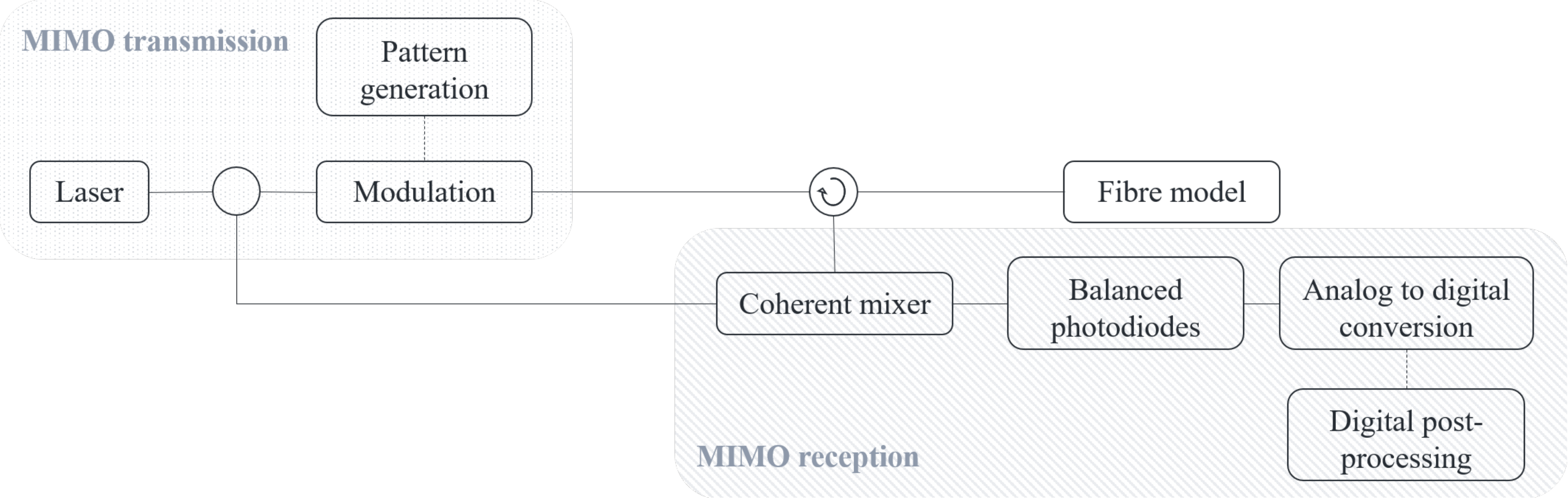}\caption{Simulated building blocks of the sensing system}\label{fig:buildingblocks}
\vspace{-7pt}
\end{figure}

\section{Phase estimation in polarization sensitive sensing}\label{sec:phaseest} 
This section focuses on the phase estimation methods to be used according to the number of used polarization states at the transmitter and receiver side.

The following notations are adopted: SISO stands for Single polarization Input - Single polarization Output sensing, which means that the light sent to the fibre sensor is linearly polarized in one direction and that the backscattered light is recovered on one polarization at the receiver side. 
SIMO stands for Single polarization Input - Multiple polarization Output sensing, which is often referred to as ``polarization diversity receiving in $\phi$-OTDR'': linearly polarized light is sent to the sensor, typically with a narrow linewidth laser, backscattered light is recovered on both polarizations using a polarization diversity receiver. 
Finally, MIMO stands for Multiple polarization Input - Multiple polarization Output sensing, where light is sent on two orthogonal polarization states at the transmitter side, and polarization diversity detection is used at the receiver. 
MISO setup is also considered: Multiple polarization Input - Single polarization Output sensing, where polarization diversity is used at the transmitter but not at the receiver.  

The Jones formalism is used. The dual-pass fibre channel is noted $\textbf{H}$,
or $\textbf{H}_i$ considering the particular segment $i$. 
We define $\textbf{H} = [\begin{smallmatrix}
h_{xx} && h_{xy} \\ h_{yx} && h_{yy}
\end{smallmatrix}$], with complex scalar coefficients.  

\subsection{Phase extraction}\label{subsec:phaseest}
MIMO allows to recover the full Jones matrix $\textbf{H}$ of the channel. 
Assuming no attenuation $A_i$ and assuming $\textbf{H}$ has a common backscattered phase $\phi$ such that 
 $\textbf{H} = e^{j\phi}[\begin{smallmatrix}
h_{11} && h_{12}\\h_{21} && h_{22} \end{smallmatrix}] $, where  
$[\begin{smallmatrix} h_{11} && h_{12}\\h_{21} && h_{22} \end{smallmatrix}] $ 
is unitary.
With $\textbf{H}$ being the dual-pass matrix of the channel, $\phi$ is the dual-pass phase. 
Hence the use of the following phase estimator:
 \begin{equation} \hat{\phi}_{MIMO} = 0.5 \angle(\det{\textbf{H}}) \label{eq:mimophase} \end{equation} 
where $\det{\textbf{H}}$ is the determinant of $\textbf{H}$. 

SIMO probes a single polarization.
For a recovered signal on both polarizations, if the first row of \textbf{H} is sent: $\textbf{H'} = [\begin{smallmatrix} h_{xx} && h_{xy} \end{smallmatrix}]^T$,
\begin{equation}
\hat{\phi}_{SIMO} = \angle(h_{xx} + h_{yx}) \label{eq:somme}
\end{equation}
Note that in Eq.~\eqref{eq:somme} we also have $\hat{\phi}_{SIMO}= \arctan(\dfrac{\Im_{xx} + \Im_{yx}}{\Re_{xx}+\Re_{yx}})$ with $\Re$, $\Im$ the real and imaginary parts of $h_{xx}$ and $h_{yx}$ data \cite{yan_coherent_2017}.

SISO recovers information on one polarization only ($h_{xx}$ for example), such that the estimated phase is $\hat{\phi}_ {SISO} = \angle(h_{xx})$. MISO proceeds the same way, and recovers $h_{xx}+h_{xy}$ with no additional information to differentiate both components. Estimated MISO phase is then $\hat{\phi}_{MISO}= \angle(h_{xx}+h_{xy})$.

Existing SISO and MISO configurations do not necessarily include a coherent receiver with polarization diversity but rather a single polarization or single $90^{\circ}$ hybrid direct detection receiver, and use all-optical techniques to recover the phase such as delay lines, dual pulses, chirped pulses...
Using direct detection, there is no sensitivity to SOP variations of the incoming light: the full intensity of the signal is received\cite{kikuchi2015fundamentals}. 
Moreover, in direct detection and coherent source scheme, it is possible to have a rather stable $\phi$-OTDR trace from comparisons of successive power-versus-time traces. 
Further trace-averaging can be performed if a speckle pattern is still noticeable \cite{goldman_direct_2013}.
Also, it was noticed that considering polarizations separately could bring a better accuracy to the system \cite{juarez_polarization_2005}.

Here, coherent detection - coherent $\phi$-OTDR is used. Coherent receivers are sensitive to the SOP of incoming light. 
To recover the full signal, there is a need to use a polarization diversity scheme, where each polarization component is detected 
by projecting the incident signal over two orthogonal polarization states and beating each projection with the same local oscillator (LO) \cite{kikuchi2015fundamentals}. 
With only one polarization projected onto the LO laser, a part of incoming light can be lost. Our SISO case refers to the latter situation. 

\subsection{Common fading phenomenons}\label{subsec:commonFading}
Several fading types were mentioned in the introduction. 
Now that the phase estimators are defined, this section aims at linking phase estimators to their respective fading risks. 

Interference fading, or coherent fading \cite{healey_fading_1984}, also referred to as Rayleigh fading and fading noise \cite{chen_high-fidelity_2018}, 
is caused by coherent interference within the same pulse due to the presence of scatterers in the fibre and to the use of a highly coherent source, resulting in a jagged intensity trace at the reception \cite{goodman_fundamental_1976}.
If a vibration occurs, the optical path changes and so does the interference figure. 
It is the reason why intensity variations due to a perturbation have no linear relationship with that perturbation. 
Such variations are
reported as an issue for OTDR techniques that rely on intensity, or that need high spatial resolution \cite{goldman_direct_2013,chen_phase-detection_2017}, since intensity variations affect locally the sensor response.  
However, coherent $\phi$-OTDR is based on that effect, 
allowing for precise phase measurement within a short time and thus over a high bandwidth, 
hence it shouldn't be mitigated using low-coherence sources  \cite{palmieri2013_distributed,goldman_direct_2013} but rather avoided. 
A way for avoiding such an effect would be to use multiple channels \cite{lin_rayleigh_2019} to probe the line. 

Polarization fading was identified in interferometric sensors \cite{stowe_polarization_1982,kersey_dependence_1988}, and remains an issue in coherent detection DAS. 
In more recent fibre sensors with coherent detection, polarization effects can prevent detection of the strength of perturbations \cite{lu_yuelan_distributed_2010}. 
Polarization fading occurs when the incident SOP and the SOP of the LO become orthogonal, which prevents mixing on the photo-detector in interferometers, and which fades the measure in heterodyne or homodyne coherent systems. 

Referring to the notations in \autoref{subsec:phaseest}, polarization fading strongly impacts SISO configuration: 
when detecting $h_{xx}$, an intensity fading will occur for periodic $(\Theta,\beta)$ pairs corresponding to orthogonality between the emitted $x$-polarization and the received $x$-polarization, resulting in 
no reliable phase estimation if the detected intensity is too low.
Conversely, SIMO and MIMO modes are supposed to be resilient to polarization fading, meaning 
there will always be a phase estimation that is stable in time: 
$\phi_{SIMO} = \angle(p_i \times (e^{j2\gamma_i} \cos2\beta_i -j\sin 2\beta_i (e^{j2\gamma_i}\cos 2\Theta_i+\sin 2\Theta_i)))$  
(here Eq.~\eqref{eq:somme} is applied to \textbf{H} with $\alpha=0$) and $\phi_{MIMO} = \angle(p_i)$. 
MISO also allows to recover an estimation as its expression is similar to that of SIMO using Eq.~\eqref{eq:somme}: even if the receiver SOP is orthogonal to polarization Y, polarization X will be recovered, and vice-versa. 
Of course, MISO and SIMO-estimated absolute phase differ from the real common phase due to polarization effects ($\Theta, \gamma$, $\beta$ appear in their expression) but when sensing dynamic events, we are only interested in phase variations compared to a stable reference.

Recent and older results, together with our definitions of SISO and SIMO, agree on the need for multiple receivers, should it be in direct detection \cite{frigo_technique_1984} or by employing polarization diversity system in coherent OTDR \cite{yang_guangyao_polarization_2016,ren_theoretical_2016}. 
In fibre interferometers, using depolarized light at the transmitter limits the polarization fading effects \cite{Kersey_1990_Analysis}. 
MISO case is not considered in the literature. 
The polarization beat length was also mentioned, as it could play a role in the quality of the received signal when confronted to polarization fading \cite{masoudi_analysis_2017}. 
This parameter has to be considered in future works.
The interest of the mentioned solutions against fading is discussed below, and comments are given on the remaining fading types along the sensor.

\section{Impact of polarization diversity on phase estimation }\label{sec:multiple_comp}\label{sec:simu}
With phase OTDR, a local phase information is processed as a differentiation between the phase estimated at one fibre segment location and that estimated at one previous location. The first fibre segment serves as a reference. 
Repeating this phase estimation process in time allows to capture and localize mechanical perturbations that may affect the fibre.
The phase retrieval is detailed below. 

In static mode, for two segments $i$ and $j$ with $j>i$, the measured differential phase is constant over time and given by $\phi_{static}=\varphi_j-\varphi_i$. 
In dynamic mode, assuming a strain is applied locally at segment position $i$, the phase is linearly disturbed relatively to the perturbation, 
following: 
\begin{equation}
\Delta \varphi_{single-pass} =  n \xi \delta l(t)\dfrac{2\pi}{\lambda} \label{eq:dphi}
\end{equation}
where $n$ is the refraction index, $\xi$ photo-elasticity coefficient ($=0.79$ for silica), $\delta l$ the fibre strain. 
So if $i$ is disturbed, the measured phase is $\phi_{dyn}=\varphi_j-\varphi_i + 2\Delta \varphi_{single-pass} = \phi_{static} + \Delta \varphi_{dual-pass} $. 
The phase delay is applied twice since the light does a round-trip in the segment (way faster than the mechanical disturbance action, so the measurement of mechanical perturbations is considered instantaneous).

For the detection of an event, we chose to monitor the standard deviation of the differential phase over time, along the fibre distance (denoted StDv). 
When an event occurs along the sensor, phase variations appear in time, thus increasing the phase standard deviation locally. 
The displayed traces in this section are simulations of a coherent detection - coherent $\phi$-OTDR, where 
the phase is recovered according to the estimators defined in \autoref{subsec:phaseest}. 
All simulations are static, e.g with no strain applied to the sensor. Therefore, this study focuses on the false alarm issue, not onto the non-detection case: 
the main requirement for phase sensors is that no variation or ``false alarm'' occurs when the fibre is not disturbed, which is also the condition to achieve the highest sensitivity. 
A simulation study of non-detection case would require a dynamic model, which is out of the scope of this paper.

\subsection{Simulations: accuracy and false alarms }\label{subsec:falseal}
Firstly, the global behaviour of the StDv along sensor distance is considered to highlight the difference between the given phase estimators. 
Secondly, paying attention to the StDv peaks, we show how they relate to the polarization rotation and birefringence in the fibre segments. 
\begin{figure}[htbp]
\centering
\captionsetup[subfigure]{oneside,margin={0cm,0cm}}
\subfloat[One simulated fibre section, log scale, 20m spatial resolution \label{subfig:stdDistComp}]{\includegraphics[width=0.333\textwidth]{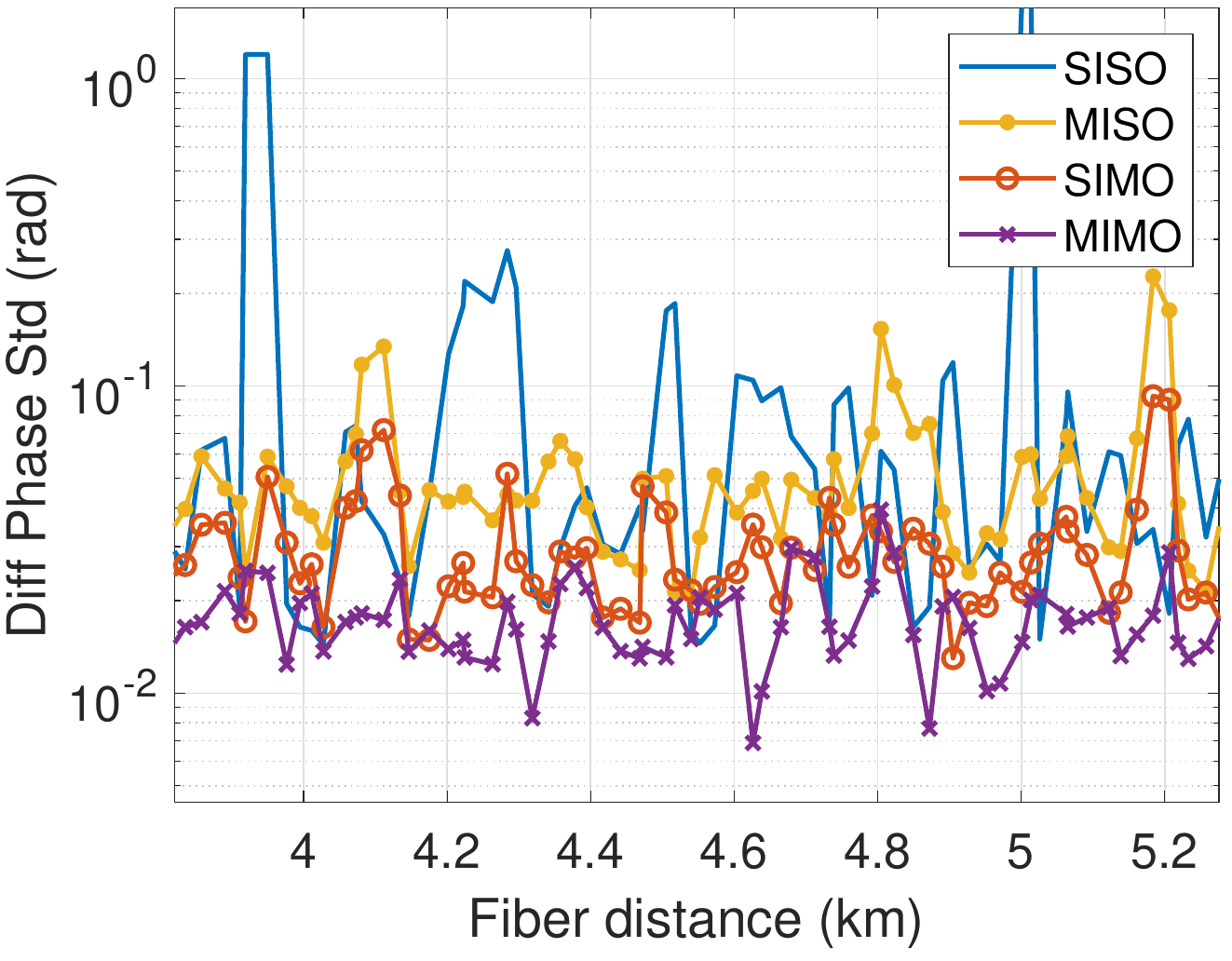}} \hspace*{0.1cm}
\subfloat[Mean StDv over 50 fibre simulations, 200m spatial resolution \label{subfig:meanstd}]{\includegraphics[width=0.333\textwidth]{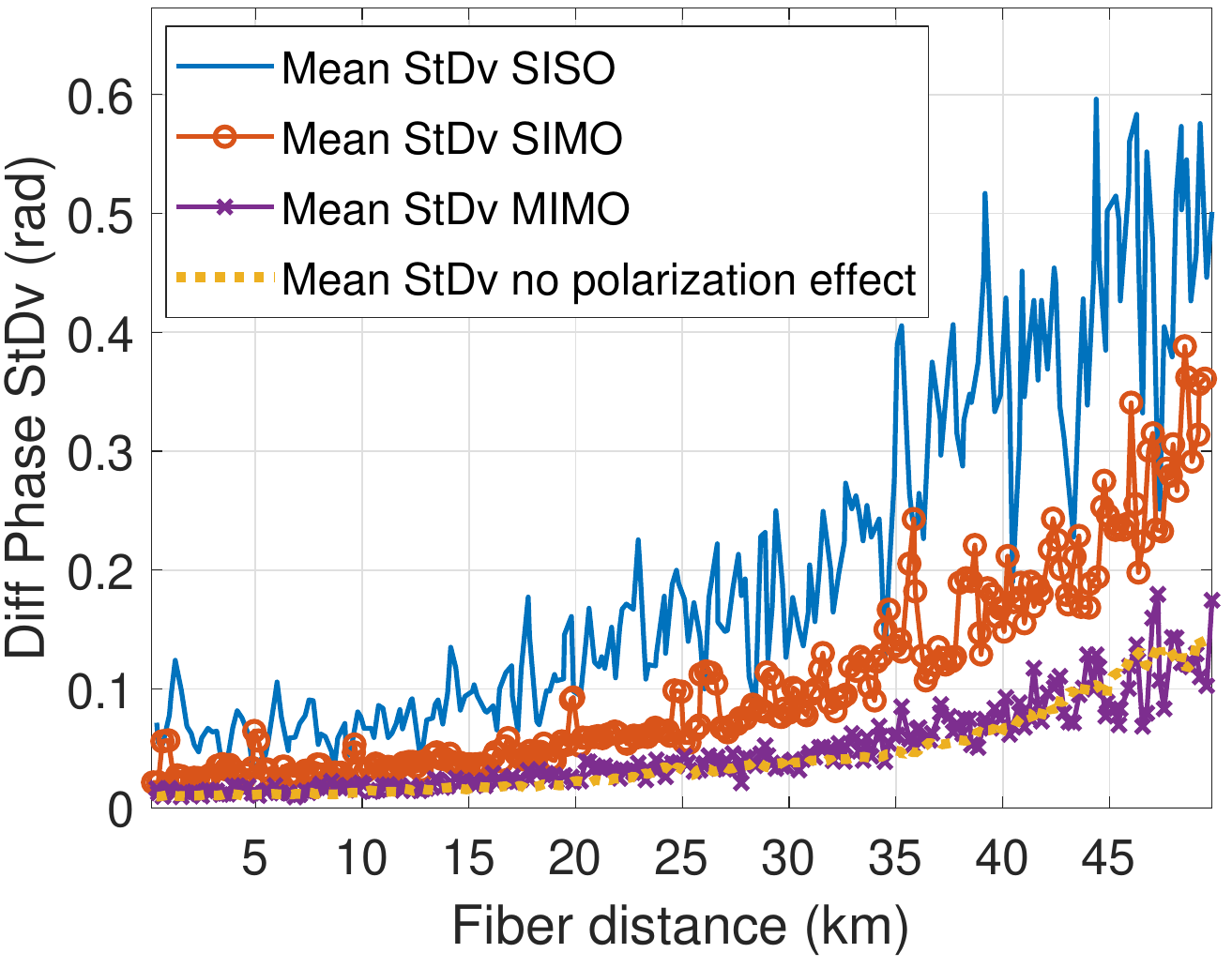}} \vspace{-7pt}
\caption{Phase StDv along distance, laser linewidth $\Delta\nu = 75\mathrm{Hz}$, 50km }
\vspace{-7pt}
\label{fig:EstMethodComp}
\end{figure}

\autoref{fig:EstMethodComp} gives the phase StDv in time along fibre distance for static simulations. 
The coded sequence duration is $5.24$ms at a symbol rate $f_{symb} = 25$Mbaud, with a constant power of $7$dBm at the entrance of the fibre sensor and $\sigma_{AWGN} = 1.7\mathrm{mV}_{RMS}$ at the receiver. The native gauge length is 4m. 
\autoref{fig:EstMethodComp}\subref{subfig:stdDistComp} is a zoom on a 1km section with 4 different phase estimators (SISO, SIMO, MISO, MIMO) for the same simulated fibre, with a spatial step of 20m ; Fig.~\ref{fig:EstMethodComp}\subref{subfig:meanstd} displays the phase StDv averaged after 50 independent 50km long fibre simulations for SISO, SIMO and MIMO phase estimators, and for a theoretical case with no polarization effects along the fibre with a spatial step of 200m.
This theoretical case is derived by taking only coherent noise into account, i.e. $\textbf{U}_i$ in Eq.~\eqref{eq:Ui} becomes the identity matrix for all fibre segments. 
\autoref{fig:EstMethodComp} shows a clear hierarchy between no polarization diversity (DP) at all, DP at one side, and DP at both TX (transmitter) and RX (receiver) sides. 
The MISO case is not displayed in \autoref{fig:EstMethodComp}\subref{subfig:meanstd} since it is comparable to SIMO.
As depicted in \autoref{fig:EstMethodComp}\subref{subfig:meanstd}, the general trend of the phase StDv over long distance is to grow exponentially with distance, with a slope depending on the laser linewidth $\Delta\nu$ \cite{Awwad2020_JLT,Guerrier2019_Model}. 
The value $\Delta\nu = 75\mathrm{Hz}$ is used in the simulations since it corresponds to the laser used in the experiments in next section \cite{Guerrier2019_Model}. 
What differentiates the phase StDv traces of the different phase estimators is the amount and dynamic of StDv peaks, or false alarms,  as shown in the single fibre simulation from \autoref{fig:EstMethodComp}\subref{subfig:stdDistComp} (e.g SISO StDv peaks are almost two orders of magnitude higher than MIMO StDv peaks). 
Deriving the mean value over several simulations as displayed in \autoref{fig:EstMethodComp}\subref{subfig:meanstd} shows how variable the phase estimation tends to be according to the chosen estimator. 
We also demonstrate in \autoref{fig:EstMethodComp}\subref{subfig:meanstd} that considering no polarization effects or using MIMO probing will lead to the same phase StDv values. 
On average, MIMO is matching the theoretical fading-free phase StDv profile: it is polarization-independent. 

What comes out is that MIMO estimated phases are the most stable with no StDv value above $5.10^{-2}$ rad over the first 5km, while SISO can vary a lot locally and quite often over fibre distance with 6 peaks above $10^{-1}$ rad within one km in \autoref{fig:EstMethodComp}\subref{subfig:stdDistComp} for example. 
SIMO and MISO show an intermediate behaviour with randomly distributed StDv peaks, here 1 to 3 StDv peaks above $10^{-1}$ rad within 1km.  
It was shown in \cite{Guerrier2020_towards} that SISO performance varies with the input as well as with the output polarization state when probing the line. 
\autoref{fig:EstMethodComp}\subref{subfig:meanstd} shows that after 50km of fibre sensor, the StDv decreases by a factor of 5 and 2 when going from SISO to MIMO and SIMO to MIMO respectively.
Furthermore, several previous works discussed in \autoref{subsec:commonFading} have already introduced polarization diversity at the receiver side to improve the sensing quality. 
Therefore, for sake of clarity, the rest of the study concentrates on the comparison between SIMO and MIMO cases only.

\begin{figure}[hbt]
\centering
\captionsetup[subfigure]{oneside,margin={0cm,0cm}}
\subfloat[SIMO, $\alpha=0$\label{subfig:simo_al0}]{\includegraphics[width=0.333\textwidth]{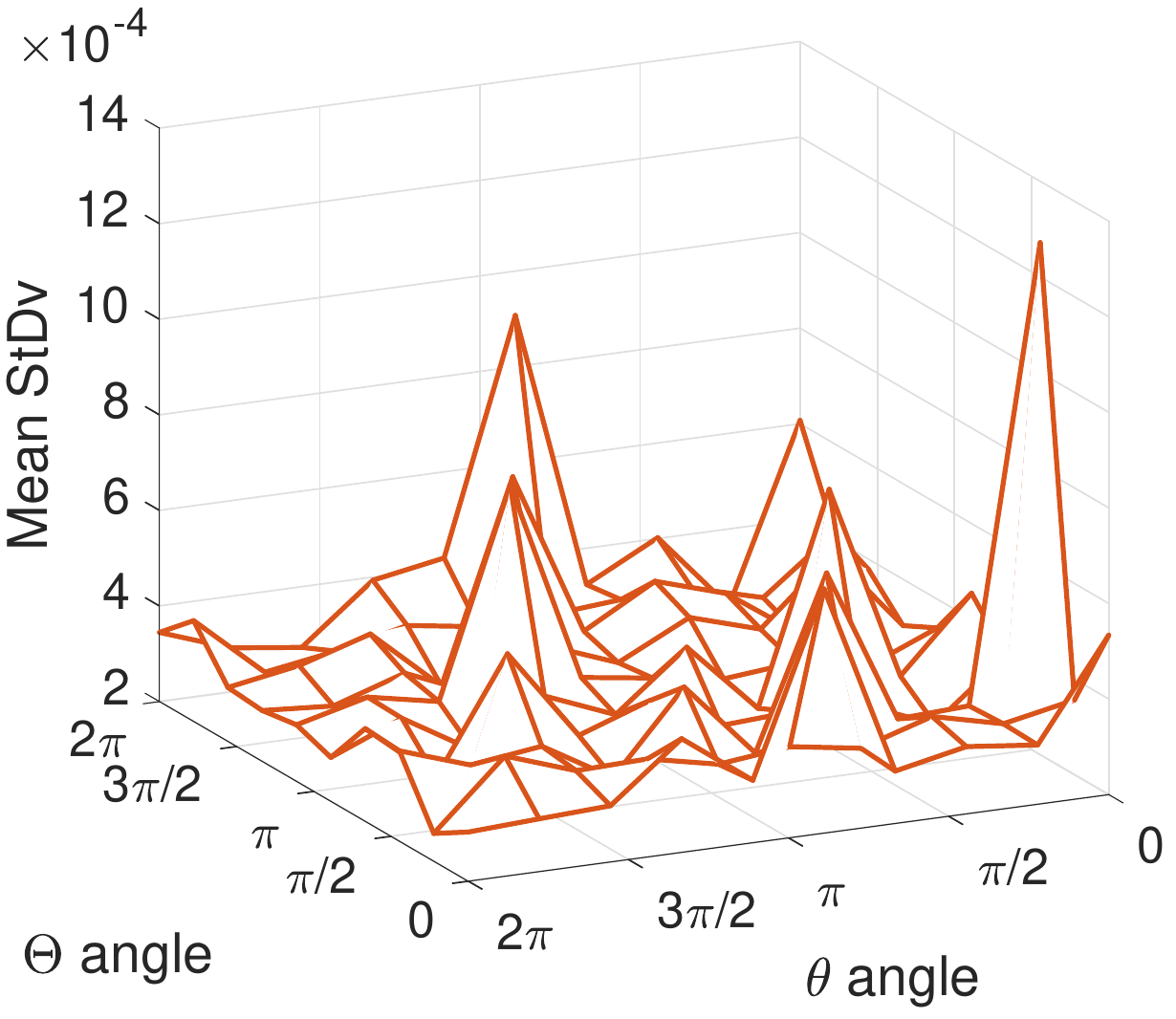}} \hspace*{1cm}
\subfloat[SIMO, $\alpha=5\%$]{\includegraphics[width=0.333\textwidth]{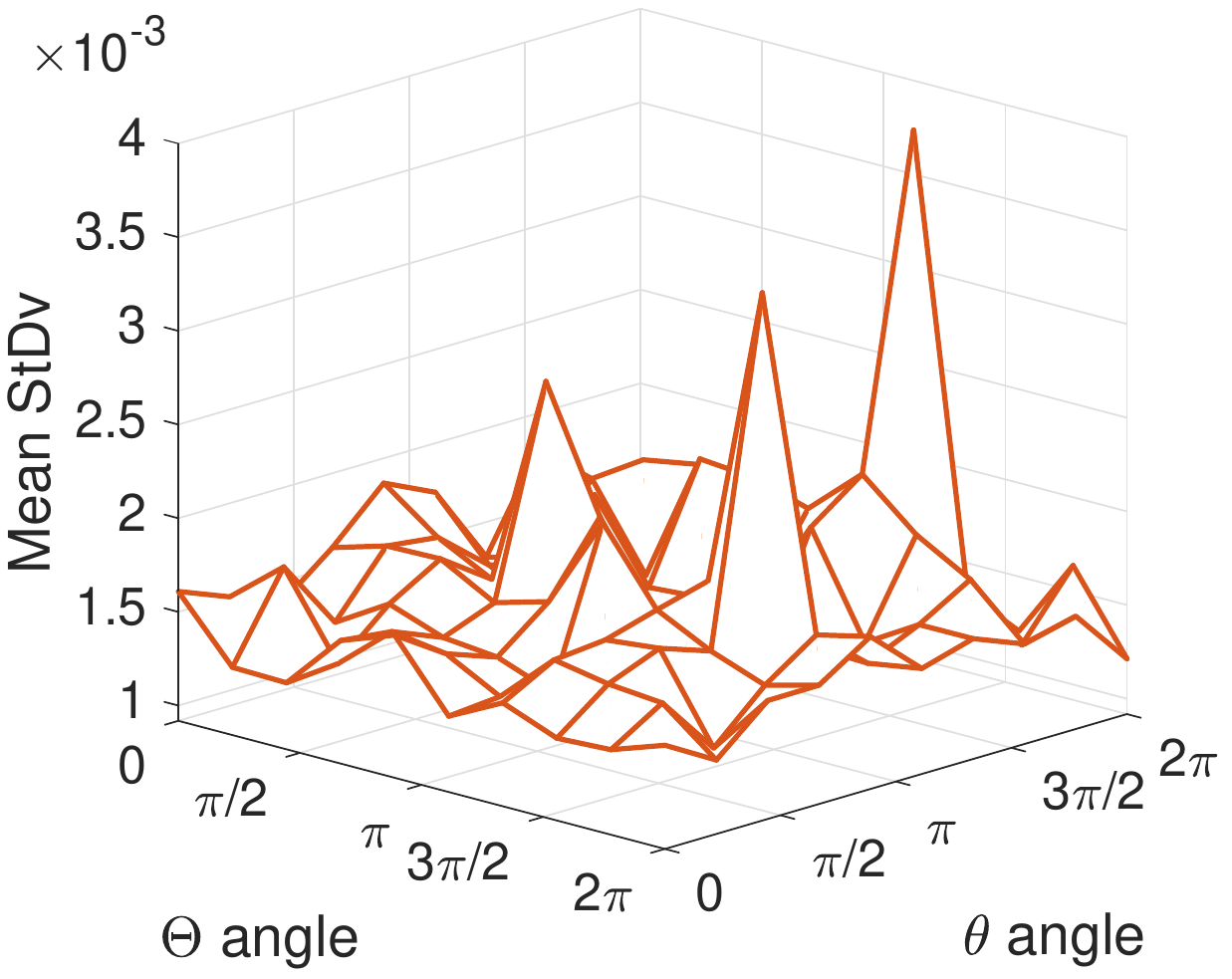}} \\
\subfloat[MIMO, $\alpha=0$]{\includegraphics[width=0.333\textwidth]{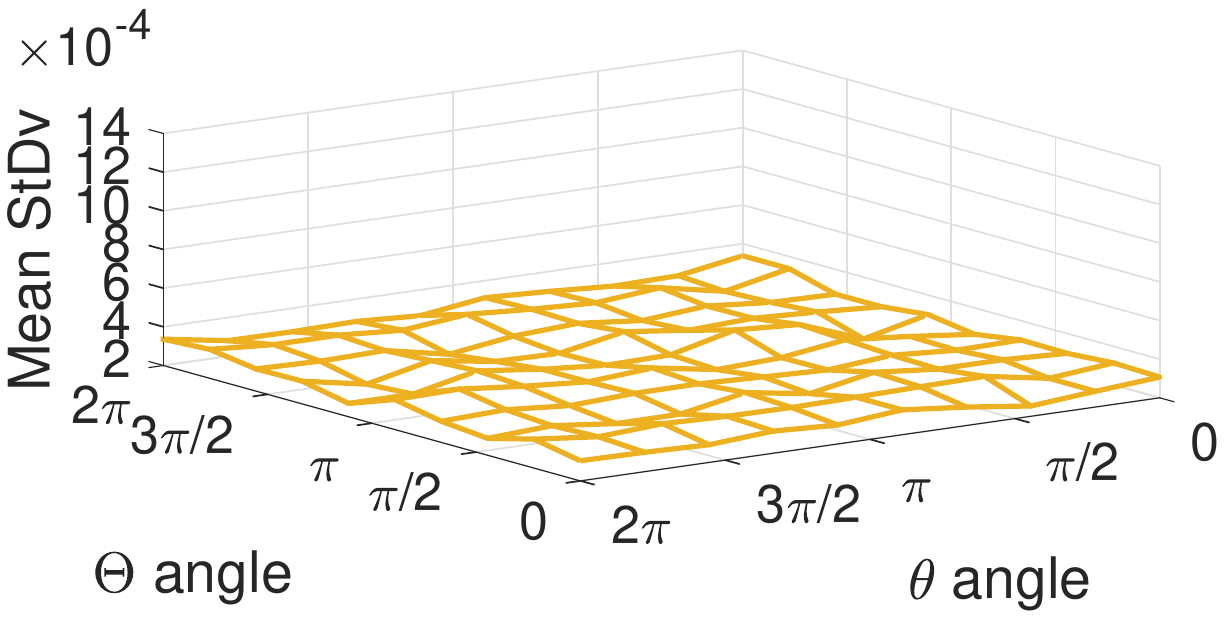}} \hspace*{1cm}
\subfloat[MIMO, $\alpha=5\%$]{\includegraphics[width=0.333\textwidth]{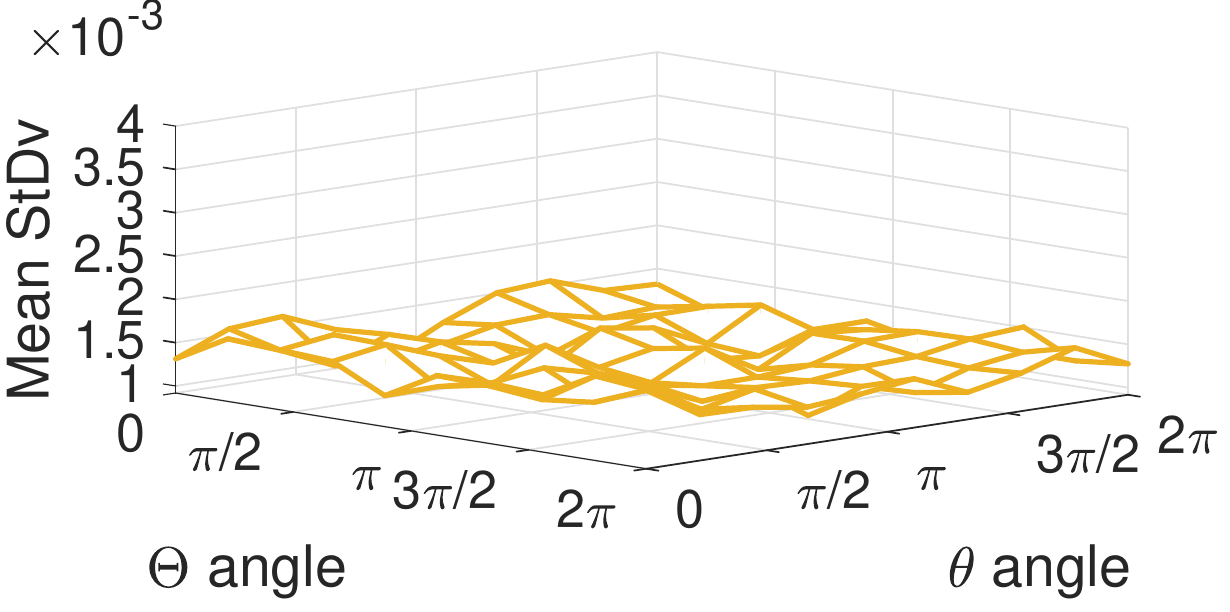}}
\caption{Phase standard deviation in one fibre segment as a function of rotations in the fibre, $\Delta\nu=75$Hz with and without $\alpha$.}\label{fig:3D-Thetatheta-Std}
\vspace{-7pt}
\end{figure}

Regardless of the phase estimation method, the evolution of the state of polarization in the fibre is described mainly by $\Theta$, $\beta$, $\gamma$ and $\theta$ defined in \autoref{subsec:fibermodel} provided that the Rayeligh scattering is an ideal reflector (no polarization crosstalk at reflection stage). 
In \autoref{fig:3D-Thetatheta-Std}\subref{subfig:simo_al0}, a 3-D view involving the evolution of angles $\Theta$ and $\theta$ for fixed randomly drawn values of $\gamma$ and $\beta$ 
together at a single segment location with a fixed reflected intensity is used to track the StDv behaviour. 
It shows that the occurence of sudden phase variations (errors on the phase estimation, or "false alarms") in SIMO is not random, but rather concentrated at some polarization parameters.  

Specific ($\Theta ,\theta$) pairs at fixed random $\gamma$ and $\beta$ 
trigger StDv peaks in SIMO, which is quite clear in Fig.~\ref{fig:3D-Thetatheta-Std}(a) and (b). On the same fibre simulations, the corresponding MIMO estimation is performed in Fig.~\ref{fig:3D-Thetatheta-Std}(c) and (d): the estimated phase is stable for all possible ($\Theta ,\theta$) pairs. 
If $\alpha \neq 0$ where $\alpha$ is the polarization transfer coefficient defined in \autoref{subsec:fibermodel} (i.e.  $\mathbf{M}$ is not an ideal reflector and introduces some crosstalk), then the locations of StDv peaks shift as shown in \autoref{fig:3D-Thetatheta-Std}(b).
The noise floor slightly rises in MIMO compared to $\alpha = 0$. 
The laser phase noise $\Delta\nu$ impacts the height of the StDv peaks and StDv values and not the general shape (position of the StDv peaks in the ($\Theta$, $\theta$) plane), so the case $\Delta\nu = 0\mathrm{Hz}$ is not displayed. 
From \autoref{fig:3D-Thetatheta-Std}, we may suspect the phase estimation to be modulated by some ($\Theta ,\theta$) "pathological" pairs which degrade the backscattered information. 
This assumption is further investigated below.

Back to \autoref{subsec:phaseest}, the SIMO phase estimator was defined as: 
$\phi_{SIMO} = \angle(p_i \times (e^{j2\gamma_i} \cos2\beta_i -j\sin 2\beta_i (e^{j2\gamma_i}\cos 2\Theta_i+\sin 2\Theta_i)))$ 
if we dismiss $\alpha$ and $\theta$.
It is recalled that $\Theta$ stands for the random polarization rotation in the fibre segment, and $\beta, \gamma$ are phase retarder coefficients, defined in \autoref{subsec:fibermodel}. 
This estimator is declared immune to intensity fading issues as it allows to get a phase estimation at all locations in the fibre, regardless of the state of polarization.

\begin{figure}[bht]
\centering
\captionsetup[subfigure]{oneside,margin={0cm,0cm}}
\subfloat[Function of $\Theta$ and $\beta$, random constant $\gamma$ between $-\pi$ and $\pi$, $\alpha=0$\label{subfig:alphazero}]{\includegraphics[width=0.39\textwidth]{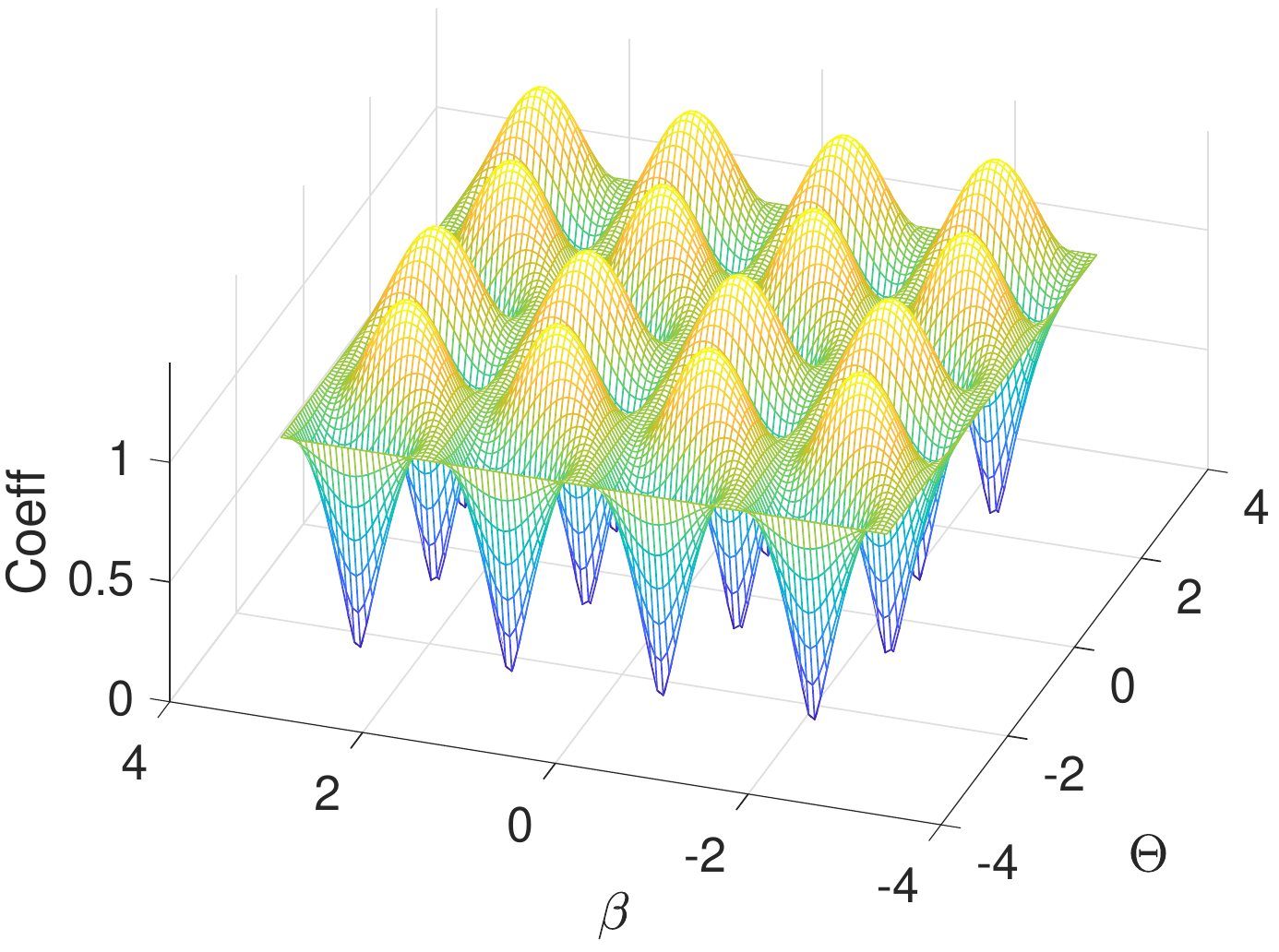}} \hspace*{1cm}
\subfloat[Function of $\Theta$ and $\beta$, random constant $\gamma$ between $-\pi$ and $\pi$, $\alpha=0.15$ \label{subfig:gamma_example}]{\includegraphics[width=0.4\textwidth]{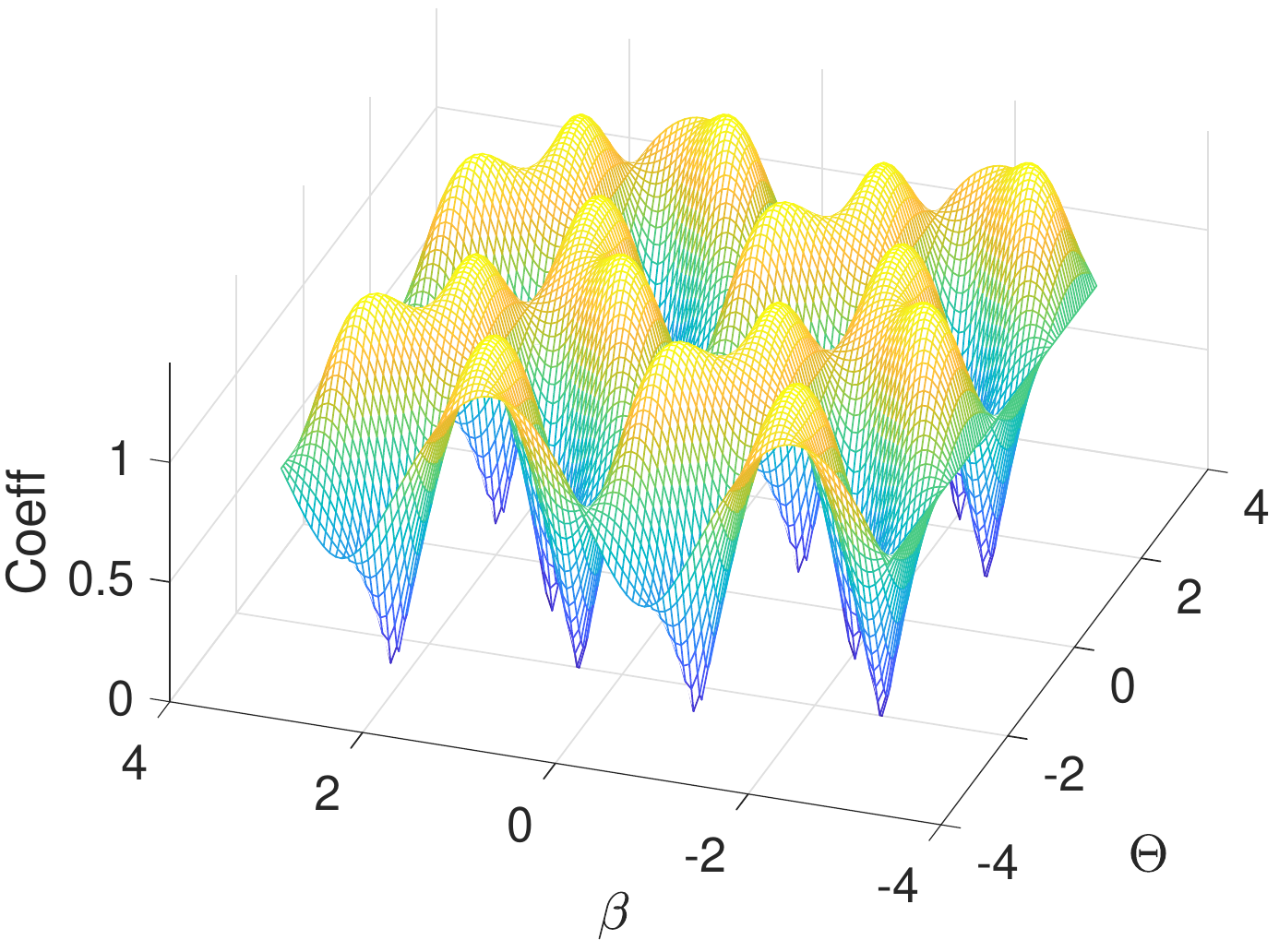}} 
\caption{Phase-fading coefficient $c_{mult,\mathrm{SIMO}}$ as function of polarization rotation parameters : SIMO phase modulation for a single fibre segment}\label{fig:thetavsgamma}
\vspace{-7pt}
\end{figure}

However, the phase $\angle{p}$ is modulated by the coefficient 
$c_{mult,\mathrm{SIMO}} = e^{j2\gamma} \cos2\beta -j\sin 2\beta \times (e^{j2\gamma} \cos 2\Theta+\sin 2\Theta)$  
This coefficient is plotted in \autoref{fig:thetavsgamma}\subref{subfig:alphazero} (for a given input angle $\theta$ and a randomly chosen $\gamma$). 
Some $(\Theta, \beta)$ pairs lead to extremely low values for that coefficient, 
which highly attenuate or "fade" the received phase value. 
For $\alpha \neq 0$, 
this phase-fading coefficient $c_{mult,\mathrm{SIMO}}$ becomes a function of $\gamma, \beta, \Theta, \alpha$ and produces a similar, still less regular pattern than in \autoref{fig:thetavsgamma}\subref{subfig:alphazero}. 
This is illustrated in \autoref{fig:thetavsgamma}\subref{subfig:gamma_example}, where a constant $\gamma$ is chosen in $[-\pi, \pi]$, and $\alpha$ is set to $0.15$. 
This configuration also has fading combinations of $\gamma, \beta, \Theta$, shown here for $\Theta , \beta$ varying in $[-\pi, \pi]$. 
It is possible to predict the values of the polarization parameters for which they cause such phase-fading.

Note that there is no phase fading with MIMO phase estimation since the equivalent coefficient for MIMO is $c_{mult,\mathrm{MIMO}} = \mathrm{abs}(\det{\textbf{H}}) = 1$ since $\textbf{H}$ is a unitary $2\times 2$ matrix.

\subsection{Expectations on sensing quality}\label{subsec:expecqua}

To begin quantifying the performance increase brought by MIMO probing and its associated phase estimation compared to SIMO, a study is conducted by means of simulations. 
Here, the fibres are probed using complementary Golay sequences on either one or two polarizations. The use of these complementary sequences allows for a mathematically 
perfect estimation of the backscattered optical field, considering either a single polarization channel, or a dual-polarization one, as demonstrated in \cite{dorize_enhancing_2018}.

The statistical study is conducted on a high number of simulations ($\mathrm{N}=10^6$ simulated points, corresponding to 2000 fibre generations for 340 m length, 40 fibres for 25 km and 20 for 50 km), with $f_{symb} = 50$Mbaud, corresponding to 2m gauge length. 
MIMO and SIMO probing simulations are run on randomly generated fibres, and the differential phase StDv is derived for every fibre segment.
The polarization beat length $L_{pb}$ is selected such that $L_{pb} \leq L_s$, assuming independent polarization state between consecutive segments (see \autoref{subsec:fibermodel}).

Three different fibre lengths are simulated to investigate the influence of distance on the StDv distribution according to the phase estimation method, so to compare the relative robustness of the SIMO and MIMO estimators over distance.

\begin{figure}[hbt]
\centering
\captionsetup[subfigure]{oneside,margin={0cm,0cm}}
\subfloat[340 m\label{subfig:340m}]{\includegraphics[width=0.333\textwidth]{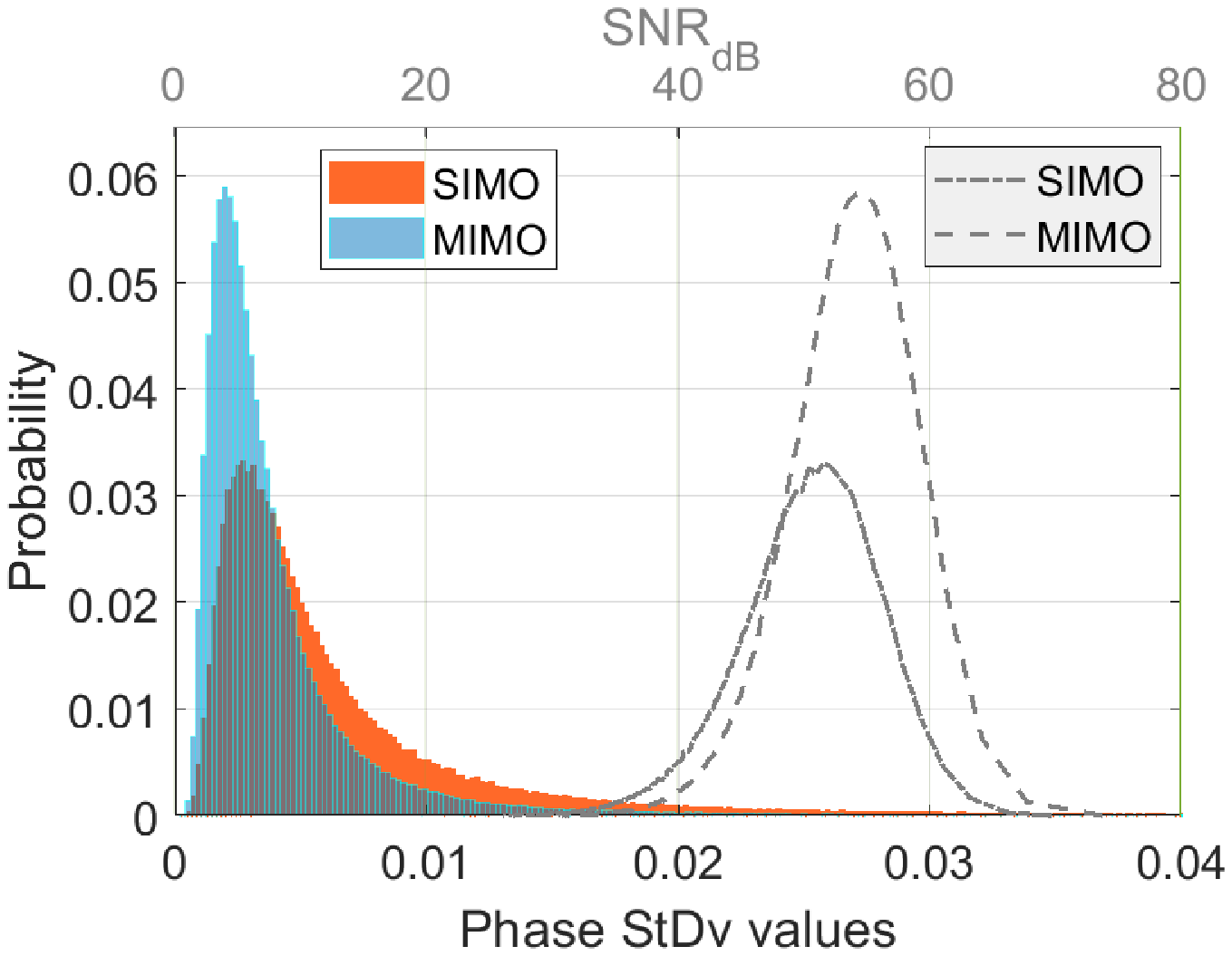}} \hspace*{0.1cm} 
\subfloat[25 km\label{subfig:25km}]{\includegraphics[width=0.333\textwidth]{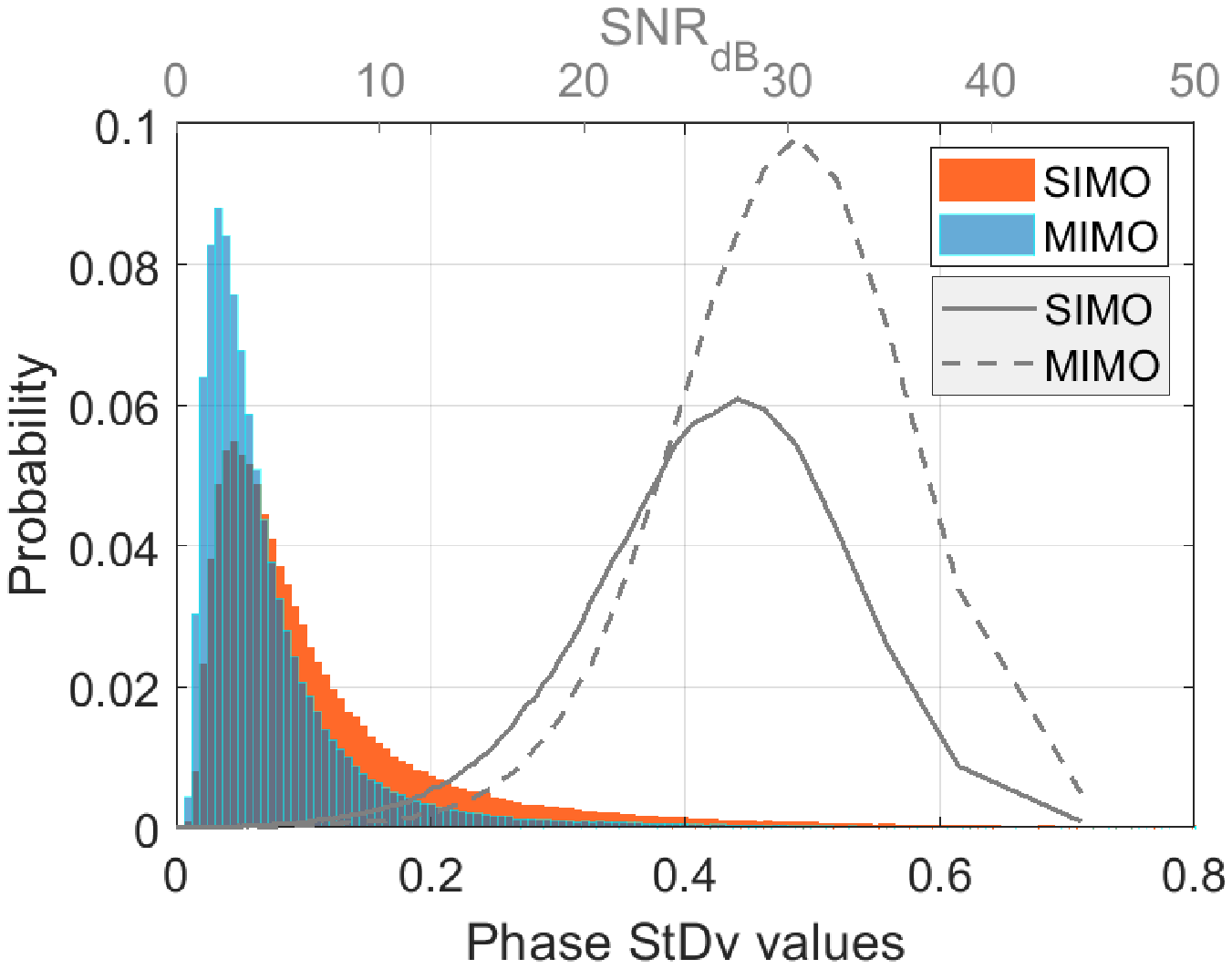}} \hspace*{0.1cm}
\subfloat[50 km\label{subfig:50km}]{\includegraphics[width=0.333\textwidth]{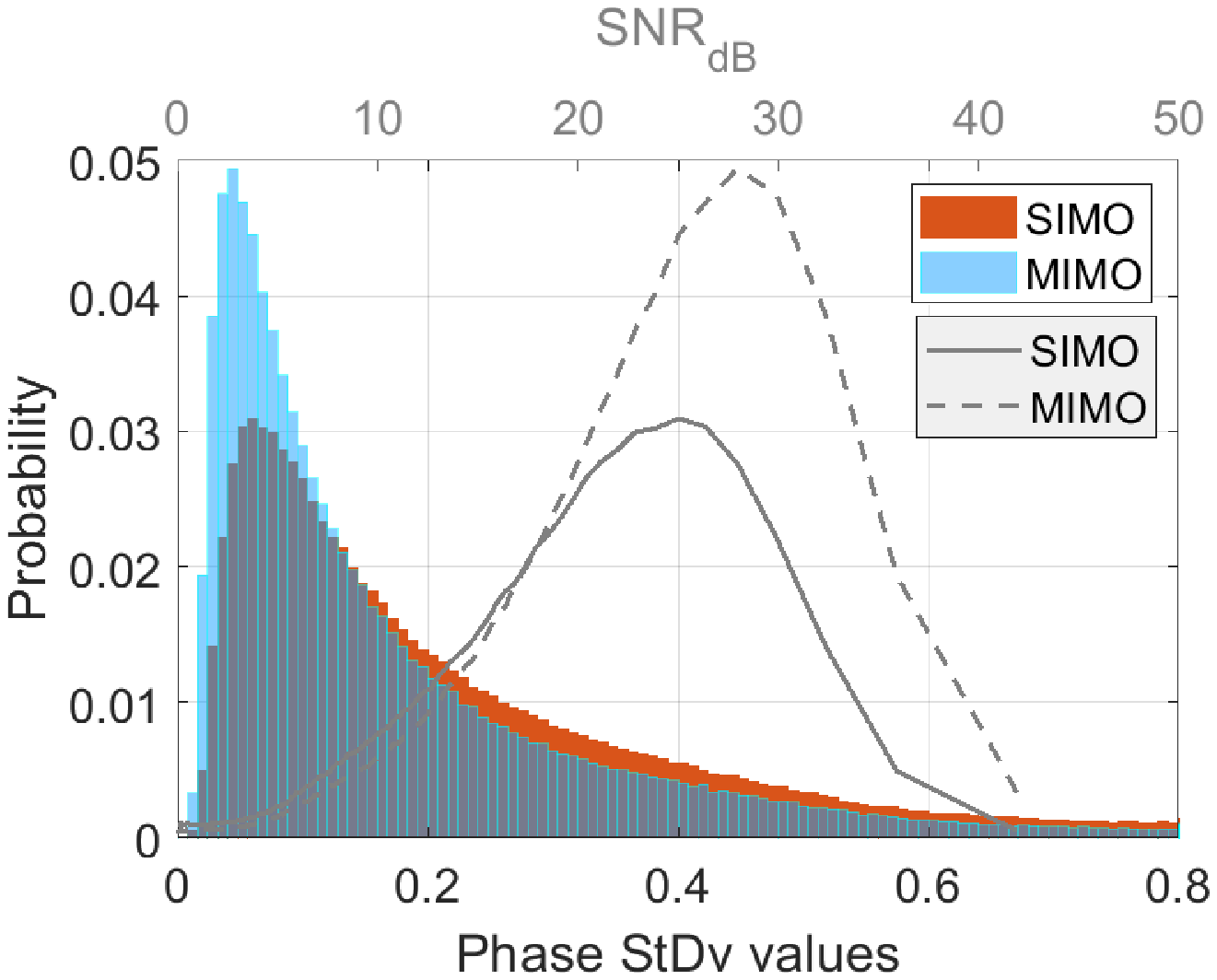}}
\caption{Distribution of differential phase StDv and corresponding $\mathrm{SNR_{phase}}$ values, SIMO and MIMO phase estimation methods, fibre simulation with $L_s= 2$m gauge length, $\Delta\nu=75$Hz, $\mathrm{N}=10^6$ simulated points}\label{fig:distrib_simomimo_length}
\vspace{-7pt}
\end{figure}

In \autoref{fig:distrib_simomimo_length}, the MIMO values of StDv are highly concentrated around low values and spread with an exponential decrease. SIMO values are less concentrated and spread further (including values $> 0.8$rad in \autoref{fig:distrib_simomimo_length}\subref{subfig:25km} and \subref{subfig:50km}). 
The distributions shape change with distance, however the same differences are observed between SIMO and MIMO cases.
We show that in \autoref{fig:distrib_simomimo_length}, between $35\%$ and $50\%$ of the SIMO estimated phases vary more than $75\%$ of MIMO estimated phases over 50km and 340m respectively, $9\%$ to $15\%$ of the SIMO estimated phases vary more than $95\%$ of MIMO estimated phases over 50km and 340m respectively.

Considering that the phase StDv is a measure of noise, we define the signal to noise ratio as $\mathrm{SNR_{phase}} = 10 \log_{10} (1/\sigma_{phase}^2)$. 
The mean value of the $\mathrm{SNR_{phase}}$ for MIMO is $1$dB above SIMO for all simulated lengths (50km, 25km, 340m). 
Moreover, the additional variance of $\mathrm{SNR_{phase}}$ in SIMO compared to MIMO method is of $\mathrm{3dBrad^2}$ at 340m, $\mathrm{11dBrad^2}$ at 25km and $\mathrm{7dBrad^2}$ at 50km. 
We conclude that MIMO enhances the mean achieved SNR and also reduces its fluctuations with respect to SIMO. 

From a high number of random simulations, it appears that SIMO phase estimations bring in more variations than MIMO, for the same simulated fibre with same scatterers distribution and birefringence, and for all distance ranges. 
The performance gap between SIMO and MIMO methods is striking for short distances, and remains significant at very long distances (50km here) when coherent fading and attenuation also affect the measurement. 
MIMO sensing appears as a strong, stable phase estimator on short fibre lengths as well as on longer distances.

\section{Experiment and sensitivity} \label{sec:expres} 
\begin{figure*}[htb]
\centering
\includegraphics[width=\textwidth]{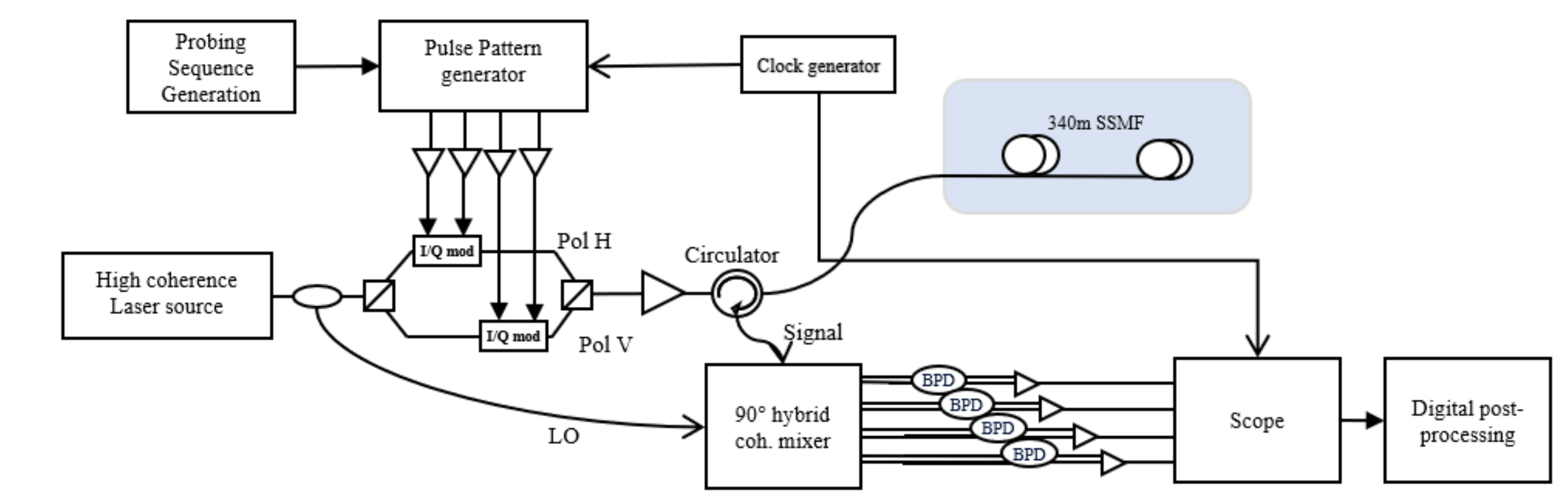}
\caption{Experimental setup. LO is Local Oscillator, BPD is Balanced PhotoDiodes}\label{fig:expsetup}
\end{figure*}
\begin{figure}[htb]
\centering
\captionsetup[subfigure]{oneside,margin={0cm,0cm}}
\subfloat[SIMO and MIMO phase StDv along fibre length\label{subfig:expfibrelength}]{\includegraphics[width=0.333\textwidth]{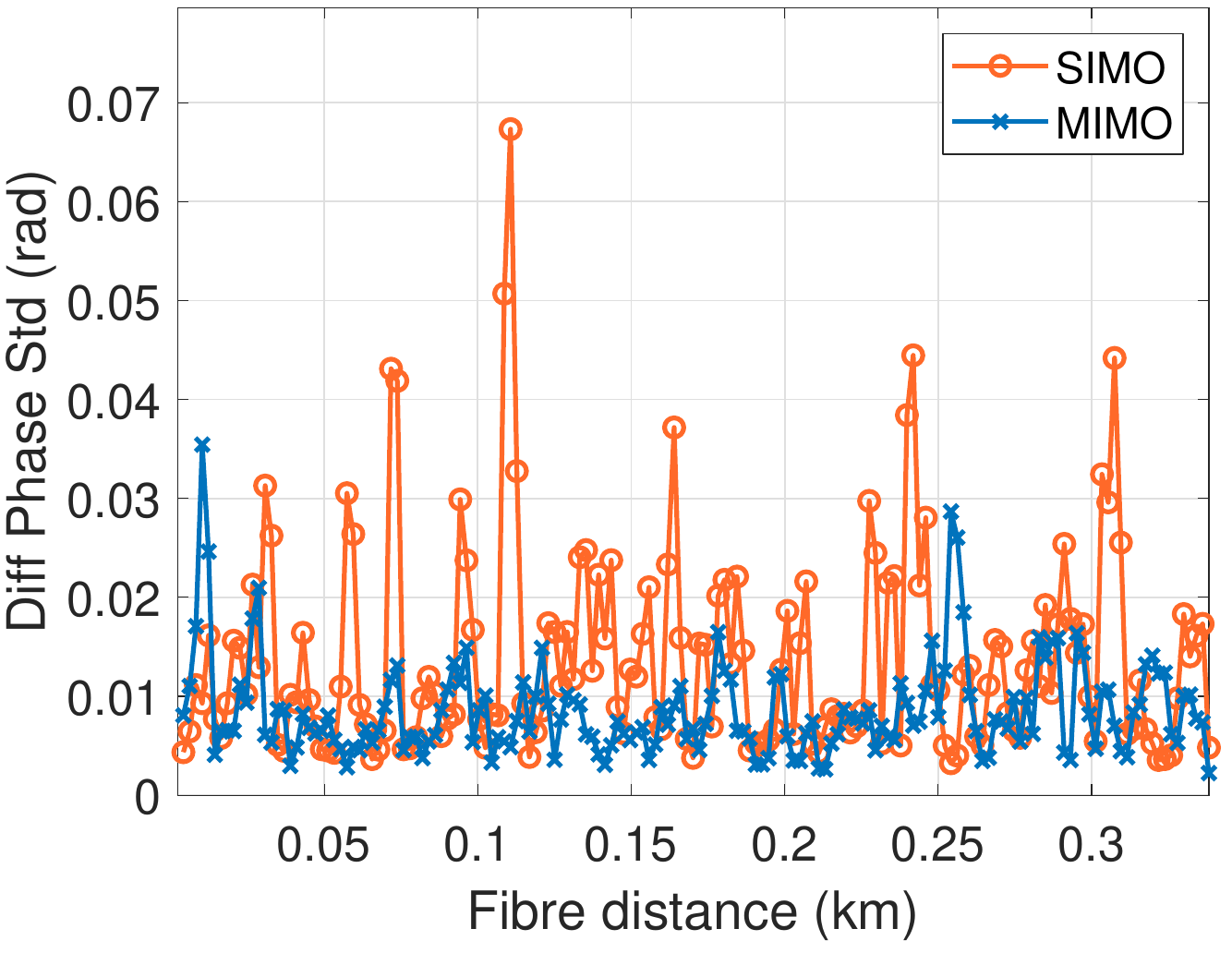}} \hspace*{1cm}
\subfloat[Distribution of SIMO and MIMO phase StDv values \label{subfig:distrib340}]{\includegraphics[width=0.35\textwidth]{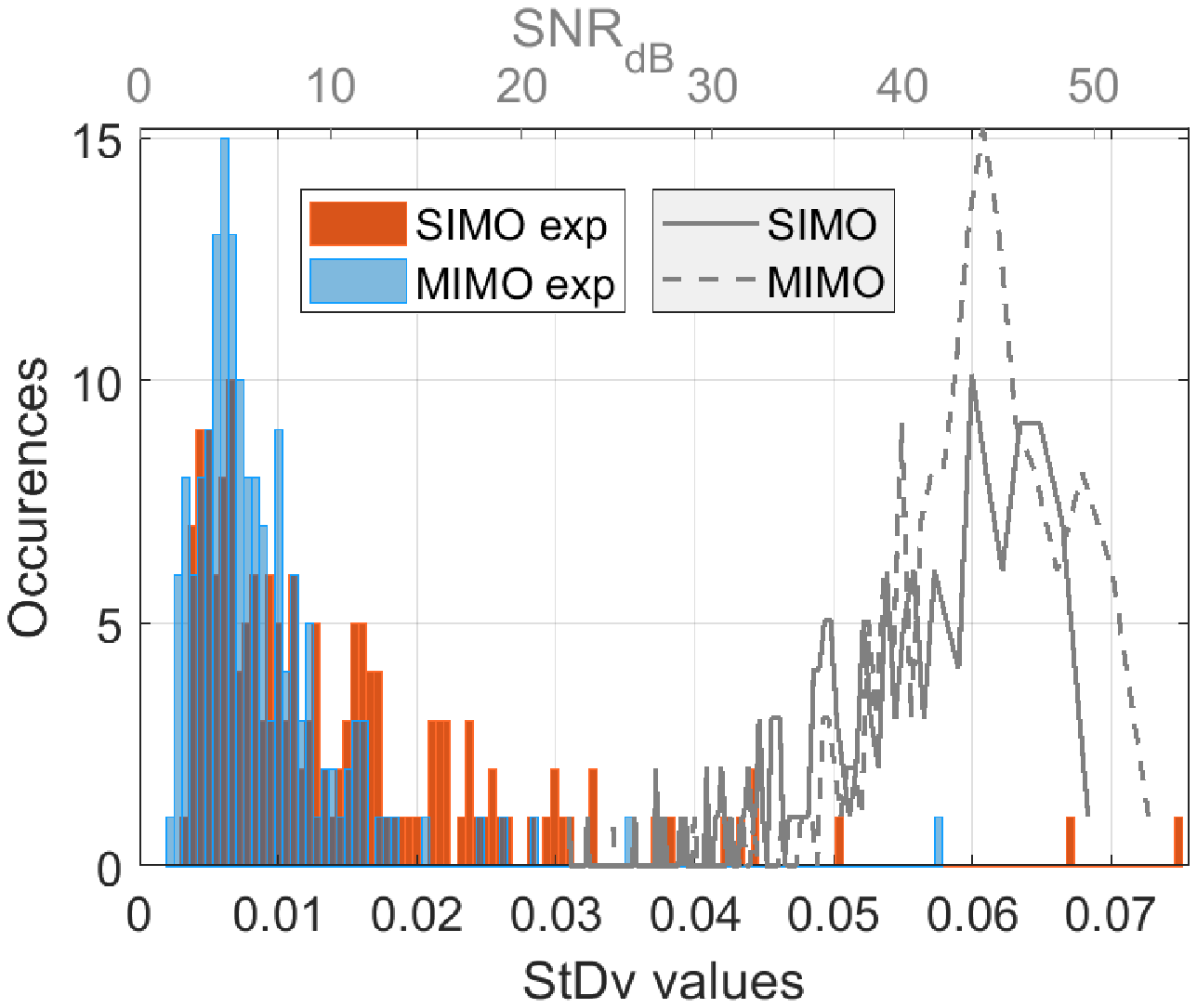}}
\caption{SIMO and MIMO measurements on 340m SSMF, no perturbation applied} \label{fig:expresult}
\vspace{-7pt}
\end{figure}

  \noindent
An experiment is performed on 340m SMF, with no perturbations around the fibre so to stay in static mode as in the simulations. 
The fibre is interrogated with BPSK sequences on one (SIMO) or both polarizations (MIMO). The setup is described in \autoref{fig:expsetup} and its operating mode further detailed in \cite{dorize_enhancing_2018}. 
SIMO and MIMO measurements could not be performed simultaneously, however they are performed successively to ensure close experimental conditions for both methods.
The mechanical bandwidth is of 1.5 kHz (50 Mbaud, 8192 symbols code length), from which the first 50 Hz bandwidth is filtered out so to avoid capturing the noise of air conditioning and other low frequency background noise sources usually met in buildings. 
The measured phase StDv, calculated from a 1s measurement window with 2m gauge length, along with its distribution are plotted in \autoref{fig:expresult}. 

Note that there is only a short distance of fibre here, so the overall measurement shows low phase standard deviation values, well below 0.1rad. 
A first look at \autoref{fig:expresult}\subref{subfig:expfibrelength} already confirms the gap of performance between SIMO and MIMO interrogation methods. 
The peaks of StDv are not perfectly aligned due to mere synchronization issues, but overall the highest StDv values for MIMO are located where the StDv is also high for SIMO. 
In these locations, the speckle pattern is suspected: coherent fading is still an issue in the fibre sensor, though it is the only one remaining with MIMO. 

The SIMO estimated phase on half of the fibre segments varies more than the MIMO estimated phase on 75\% of the (same) fibre segments. 
It is consistent with the values obtained through simulation in \autoref{subsec:expecqua}. 
As for the $\mathrm{SNR_{phase}}$ values, and despite the restricted number of measured points, the variance of $\mathrm{SNR_{phase}}$ is $2.3\mathrm{dBrad^2}$ higher in SIMO compared to MIMO, which is consistent with the simulation value ($\mathrm{3dBrad^2}$ at 340m). 
 Although no SIMO probing was experimented on longer fibre, it is expected that the relative distribution of SIMO and MIMO is not supposed to vary much. 
 In a recent work \cite{dorize_high_2019}, we demonstrated a good stability of MIMO phase estimation on several tens of km.

 Therefore, these experimental results, in line with previous simulation ones, confirm the better sensitivity gain brought by a Coherent-MIMO approach in phase OTDR. 

\section{Conclusion} \label{sec:ccl}
The exploitation of two orthogonal polarization channels in an optical fibre was highlighted in the case of optical fibre sensing. 
From a numerical model of the Rayleigh backscattering in single-mode fibres, simulations were run for different configurations of transmitter and receiver. 
Several phase estimation methods for coherent $\phi$-OTDR were compared, among which Single polarization Input - Multiple polarization Output sensing (SIMO), often referred to as ``polarization diversity receiving in $\phi$-OTDR" in the literature, and Multiple polarization Input - Multiple polarization Output sensing (MIMO). 
It was demonstrated that some sensing configurations, in particular those with no polarization diversity, were sensitive to fading effects and could neither allow to correctly recover the intensity nor the phase information from all fibre locations.
Even with polarization diversity at the receiver, we pointed out that SIMO estimated phase is subject to a modulation 
which biases the phase estimation for some fibre segments and triggers local phase variations. 
It was shown both in simulation and through experiment performed in the absence of mechanical perturbation that SIMO estimated phases vary considerably more than MIMO estimated ones, for the same fibre in the same conditions:
Coherent-MIMO, by completely mitigating the effects of polarization, appears twice more sensitive and more reliable.  

In essence, 
we were able to thoroughly compare existing interrogation methods for coherent optical fibre sensors, and so to demonstrate the superiority of Coherent-MIMO over partial polarization diversity sensing techniques in terms of sensitivity. It proves that further research on dual-polarization interrogation of fibre sensors is of great interest, and gives insights on the key concerns towards a more reliable and sensitive fibre sensing. 

\section*{Acknowledgments}
The authors would like to warmly thank Robert M. Jopson, from Bell Labs, for constructive discussions on the physical implications of the dual-polarization model.

\section*{Disclosures}
The authors declare no conflicts of interest.

\bibliography{bibliographie}

\begin{thebibliography}{10}
\newcommand{\enquote}[1]{``#1''}

\bibitem{Hartog2018_advances}
A.~H. Hartog, M.~Belal, and M.~A. Clare, \enquote{Advances in distributed
  fiber-optic sensing for monitoring marine infrastructure, measuring the deep
  ocean, and quantifying the risks posed by seafloor hazards,}
  {\protect\JournalTitle{Marine Technology Society Journal}} \textbf{52},
  58--73 (2018).

\bibitem{beisenova_fiber-optic_2018}
A.~Beisenova, A.~Issatayeva, D.~Tosi, and C.~Molardi, \enquote{Fiber-optic
  distributed strain sensing needle for real-time guidance in epidural
  anesthesia,} {\protect\JournalTitle{{IEEE} Sensors Journal}} \textbf{18},
  8034--8044 (2018).

\bibitem{palmieri2013_distributed}
L.~Palmieri and L.~Schenato, \enquote{Distributed optical fiber sensing based
  on rayleigh scattering,} {\protect\JournalTitle{The Open Optics Journal}}
  \textbf{7} (2013).

\bibitem{pastor_2016_temperature}
J.~Pastor-Graells, H.~Martins, A.~Garcia-Ruiz, S.~Martin-Lopez, and
  M.~Gonzalez-Herraez, \enquote{Single-shot distributed temperature and strain
  tracking using direct detection phase-sensitive {OTDR} with chirped pulses,}
  {\protect\JournalTitle{Optics express}} \textbf{24}, 13121--13133 (2016).

\bibitem{Lu_2017_highSpaRes}
B.~Lu, Z.~Pan, Z.~Wang, H.~Zheng, Q.~Ye, R.~Qu, and H.~Cai, \enquote{High
  spatial resolution phase-sensitive optical time domain reflectometer with a
  frequency-swept pulse,} {\protect\JournalTitle{Opt. Lett.}} \textbf{42},
  391--394 (2017).

\bibitem{dorize_enhancing_2018}
C.~Dorize and E.~Awwad, \enquote{Enhancing the performance of coherent {OTDR}
  systems with polarization diversity complementary codes,}
  {\protect\JournalTitle{Opt. Express}} \textbf{26}, 12878--12890 (2018).

\bibitem{SEAFOM_2018}
SEAFOM, \emph{DAS Parameter Definitions and Tests}, http://seafom.com,
  measuring sensor performance-02 ed. (2018).

\bibitem{Deventer1993}
M.~O.~V. Deventer, \enquote{Polarization properties of rayleigh backscattering
  in single-mode fibers,} {\protect\JournalTitle{Journal of Lightwave
  Technology}} \textbf{11}, 1895--1899 (1993).

\bibitem{goodman_fundamental_1976}
J.~W. Goodman, \enquote{Some fundamental properties of speckle*,}
  {\protect\JournalTitle{Journal of the Optical Society of America}}
  \textbf{66}, 1145 (1976).

\bibitem{healey_fading_1984}
P.~Healey, \enquote{Fading in heterodyne {OTDR},}
  {\protect\JournalTitle{Electronics Letters}} \textbf{20}, 30--32 (1984).

\bibitem{Fleyer2015}
M.~Fleyer, J.~P. Cahill, M.~Horowitz, C.~R. Menyuk, and O.~Okusaga,
  \enquote{Comprehensive model for studying noise induced by self-homodyne
  detection of backward rayleigh scattering in optical fibers,}
  {\protect\JournalTitle{Opt. Express}} \textbf{23}, 25635--25652 (2015).

\bibitem{chen_phase-detection_2017}
D.~Chen, Q.~Liu, and Z.~He, \enquote{Phase-detection distributed fiber-optic
  vibration sensor without fading-noise based on time-gated digital {OFDR},}
  {\protect\JournalTitle{Optics Express}} \textbf{25}, 8315 (2017).

\bibitem{yan_coherent_2017}
Q.~Yan, M.~Tian, X.~Li, Q.~Yang, and Y.~Xu, \enquote{Coherent $\phi$-{OTDR}
  based on polarization-diversity integrated coherent receiver and heterodyne
  detection,} in \emph{2017 25th Optical Fiber Sensors Conference ({OFS}),}
  (2017), pp. 1--4.

\bibitem{martins_real_2016}
H.~F. Martins, K.~Shi, B.~C. Thomsen, S.~Martin-Lopez, M.~Gonzalez-Herraez, and
  S.~J. Savory, \enquote{Real time dynamic strain monitoring of optical links
  using the backreflection of live {PSK} data,} {\protect\JournalTitle{Optics
  Express}} \textbf{24}, 22303 (2016).

\bibitem{yang_guangyao_polarization_2016}
G.~Yang, X.~Fan, B.~Wang, Q.~Liu, and Z.~He, \enquote{Polarization fading
  elimination in phase-extracted {OTDR} for distributed fiber-optic vibration
  sensing,} in \emph{2016 21st {OptoElectronics} and Communications Conference
  ({OECC}) held jointly with 2016 International Conference on Photonics in
  Switching ({PS}),}  (2016), pp. 1--3.

\bibitem{gu_comparison_2018}
F.~Gu, Y.~Li, L.~Liang, and M.~Zhang, \enquote{Comparison of three combining
  methods for polarization-diversity receiving in $\varphi$ -{OTDR},} in
  \emph{2018 23rd Opto-Electronics and Communications Conference ({OECC}),}
  (2018), pp. 1--2. {ISSN}: 2166-8884.

\bibitem{Kersey_1988_Observation}
A.~D. Kersey, M.~J. Marrone, and A.~Dandridge, \enquote{Observation of
  input-polarization-induced phase noise in interferometric fiber-optic
  sensors,} {\protect\JournalTitle{Opt. Lett.}} \textbf{13}, 847--849 (1988).

\bibitem{Kersey_1990_Analysis}
A.~D. {Kersey}, M.~J. {Marrone}, and A.~{Dandridge}, \enquote{Analysis of
  input-polarization-induced phase noise in interferometric fiber-optic sensors
  and its reduction using polarization scrambling,}
  {\protect\JournalTitle{Journal of Lightwave Technology}} \textbf{8}, 838--845
  (1990).

\bibitem{Guerrier2019_Model}
S.~Guerrier, C.~Dorize, E.~Awwad, and J.~Renaudier, \enquote{A
  dual-polarization rayleigh backscatter model for phase-sensitive {OTDR}
  applications,} {\protect\JournalTitle{Optical Sensors and Sensing
  Conference}}  (2019).

\bibitem{Liokumovich2015}
L.~B. Liokumovich, N.~A. Ushakov, O.~I. Kotov, M.~A. Bisyarin, and A.~H.
  Hartog, \enquote{Fundamentals of optical fiber sensing schemes based on
  coherent optical time domain reflectometry: Signal model under static fiber
  conditions,} {\protect\JournalTitle{Journal of Lightwave Technology}}
  \textbf{33}, 3660--3671 (2015).

\bibitem{masoudi_analysis_2017}
A.~Masoudi and T.~P. Newson, \enquote{Analysis of distributed optical fibre
  acoustic sensors through numerical modelling,} {\protect\JournalTitle{Optics
  Express}} \textbf{25}, 32021 (2017).

\bibitem{BOYA_2003_volumes}
L.~J. Boya, E.~Sudarshan, and T.~Tilma, \enquote{Volumes of compact manifolds,}
  {\protect\JournalTitle{Reports on Mathematical Physics}} \textbf{52}, 401 --
  422 (2003).

\bibitem{Damask2004_chap2}
J.~N. Damask, \enquote{Construction of general unitary matrix,} in
  \emph{Polarization optics in telecommunications,}  vol. 101 (Springer Science
  \& Business Media, 2004), chap. The spin-vector calculus of polarization, pp.
  50--71.

\bibitem{sosman1927properties}
R.~B. Sosman, \emph{The Properties of Silica: An introduction to the properties
  of substances in the solid non-conducting state}, vol.~37 (Book Department,
  The Chemical Catalog Company, Incorporated, 1927).

\bibitem{Hartog2017_chap2}
A.~H. Hartog, \enquote{Propagation in optical fibres,} in \emph{An Introduction
  to Distributed Optical Fibre Sensors,}  (CRC Press, 2017), chap. 2.1.

\bibitem{Ross1982}
J.~N. Ross, \enquote{Birefringence measurement in optical fibers by
  polarization-optical time-domain reflectometry,}
  {\protect\JournalTitle{Applied Optics}} \textbf{21}, 3489 (1982).

\bibitem{Awwad2020_JLT}
E.~Awwad, C.~Dorize, S.~Guerrier, and J.~Renaudier,
  \enquote{Detection-localization-identification of vibrations over long
  distance {SSMF} with coherent delta-phi-{OTDR},}
  {\protect\JournalTitle{Journal of Lightwave Technology}} \textbf{38},
  3089--3095 (2020).

\bibitem{corsi_beat_1999}
F.~Corsi, A.~Galtarossa, and L.~Palmieri, \enquote{Beat length characterization
  based on backscattering analysis in randomly perturbed single-mode fibers,}
  {\protect\JournalTitle{Journal of Lightwave Technology}} \textbf{17},
  1172--1178 (1999).

\bibitem{kikuchi2015fundamentals}
K.~Kikuchi, \enquote{Fundamentals of coherent optical fiber communications,}
  {\protect\JournalTitle{Journal of Lightwave Technology}} \textbf{34},
  157--179 (2015).

\bibitem{goldman_direct_2013}
R.~Goldman, A.~Agmon, and M.~Nazarathy, \enquote{Direct detection and coherent
  optical time-domain reflectometry with golay complementary codes,}
  {\protect\JournalTitle{Journal of Lightwave Technology}} \textbf{31},
  2207--2222 (2013).

\bibitem{juarez_polarization_2005}
J.~C. Juarez and H.~F. Taylor, \enquote{Polarization discrimination in a
  phase-sensitive optical time-domain reflectometer intrusion-sensor system,}
  {\protect\JournalTitle{Optics Letters}} \textbf{30}, 3284 (2005).

\bibitem{chen_high-fidelity_2018}
D.~Chen, Q.~Liu, and Z.~He, \enquote{High-fidelity distributed fiber-optic
  acoustic sensor with fading noise suppressed and sub-meter spatial
  resolution,} {\protect\JournalTitle{Optics Express}} \textbf{26}, 16138
  (2017).

\bibitem{lin_rayleigh_2019}
S.~Lin, Z.~Wang, J.~Xiong, Y.~Fu, J.~Jiang, Y.~Wu, Y.~Chen, C.~Lu, and Y.~Rao,
  \enquote{Rayleigh fading suppression in one-dimensional optical scatters,}
  {\protect\JournalTitle{{IEEE} Access}} \textbf{7}, 17125--17132 (2019).

\bibitem{stowe_polarization_1982}
D.~W. Stowe, D.~R. Moore, and R.~G. Priest, \enquote{Polarization fading in
  fiber interferometric sensors,} {\protect\JournalTitle{{IEEE} Transactions on
  Microwave Theory and Techniques}} \textbf{30}, 1632--1635 (1982).

\bibitem{kersey_dependence_1988}
A.~D. Kersey, A.~Dandridge, and A.~B. Tveten, \enquote{Dependence of visibility
  on input polarization in interferometric fiber-optic sensors,}
  {\protect\JournalTitle{Opt. Lett.}} \textbf{13}, 288--290 (1988).

\bibitem{lu_yuelan_distributed_2010}
{Lu, Yuelan}, {Tao Zhu}, {Liang Chen}, and {Xiaoyi Bao}, \enquote{Distributed
  vibration sensor based on coherent detection of phase-{OTDR},}
  {\protect\JournalTitle{Journal of Lightwave Technology}} p. 5585644 (2010).

\bibitem{frigo_technique_1984}
N.~Frigo, A.~Dandridge, and A.~Tveten, \enquote{Technique for elimination of
  polarisation fading in fibre interferometers,}
  {\protect\JournalTitle{Electronics Letters}} \textbf{20}, 319--320 (1984).

\bibitem{ren_theoretical_2016}
M.~Ren, P.~Lu, L.~Chen, and X.~Bao, \enquote{Theoretical and experimental
  analysis of $\phi$-{OTDR} based on polarization diversity detection,}
  {\protect\JournalTitle{{IEEE} Photonics Technology Letters}} \textbf{28},
  697--700 (2016).

\bibitem{Guerrier2020_towards}
S.~Guerrier, C.~Dorize, E.~Awwad, and J.~Renaudier, \enquote{Towards
  polarization-insensitive coherent coded phase {OTDR},} in \emph{2020 27th
  Optical Fiber Sensors Conference ({OFS}),}  (Optical Society of America, 2020
  accepted, to be published).

\bibitem{dorize_high_2019}
C.~Dorize, E.~Awwad, and J.~Renaudier, \enquote{High sensitivity
  $\varphi$-{OTDR} over long distance with polarization multiplexed codes,}
  {\protect\JournalTitle{{IEEE} Photonics Technology Letters}} \textbf{31},
  1654--1657 (2019).

\end{thebibliography}

\end{document}